\documentclass[11pt]{article}
\usepackage{graphicx,caption2,subfigure}
\makeatletter
\newcommand{\singlespacing}{\let\CS=\@currsize\renewcommand{\baselinestretch}{1.0}\tiny\CS}
\newcommand{\doublespacing}{\let\CS=\@currsize\renewcommand{\baselinestretch}{1.5}\tiny\CS}

\begin{document}
\title {On Production of `Soft' Particles in Au+Au and Pb+Pb Collisions at High Energies}
\author {A. C. Das
Ghosh$^1$\thanks{e-mail: dasghosh@yahoo.co.in}, Goutam
Sau$^2$\thanks{e-mail: sau$\_$goutam@yahoo.com}, S. K.
Biswas$^3$\thanks{e-mail: sunil$\_$biswas2004@yahoo.com}
 $\&$ S. Bhattacharyya$^4$\thanks{e-mail: bsubrata@www.isical.ac.in
(Communicating Author).}\\
{\small $^1$ Department of Microbiology,}\\
    {\small Surendranath College, Kolkata-700009, India.}\\
  {\small $^2$ Beramara RamChandrapur High School,}\\
 {\small South 24-Pgs, 743609(WB), India.}\\
{\small $^3$ West Kodalia Adarsha Siksha Sadan,}\\
  {\small New Barrackpore, Kolkata-700131, India.}\\
  {\small $^4$ Physics and Applied Mathematics Unit(PAMU),}\\
 {\small Indian Statistical Institute, Kolkata - 700108, India.}}
\date{}
\maketitle
\bigskip
\begin{abstract}
 Production of low-$p_T$ (soft) hadronic particles in high energy collisions
constitutes a significant corner of special interests and problems,
as the perturbative quantum chromodynamics (pQCD) does not work in
this region. We have probed here into the nature of the light
particle production in two symmetric nuclear collisions at two
neighbouring energies with the help of two non-standard models. The
results are found to be in good agreement with data. Despite this,
as the models applied here are not intended to provide deep insights
into the actual physical processes involved in such collisions, the
phenomenological bounds and constraints which cannot be remedied for
the present continue to exist.
\end{abstract}
\newpage
\doublespacing
\section{Introduction}
Multiple production of hadrons at high energies, named also
multiparticle production phenomena, is still not at all a
well-understood sector. In general, the largest bulk of the
particles produced in nature, called secondaries, is detected to be
of the small transverse momenta, though in arranged laboratory
collisions at colliders (inclusive of LHC) the particles are, in the
main, detected with and reported for large transverse momenta. The
theoretical studies based on the standard model(SM) of particle
interaction are grounded on an artificial division between
'soft'(low-$p_T$) and 'hard'(large-$p_T$) regions. The latter
wherein the perturbative techniques are applied is called
Perturbative Quantum Chromodynamics (pQCD); and the former wherein
perturbation theory fails is termed non-perturbative domain. And
most of the particles in nature as well as in laboratory collisions
fall under this non-perturbative sector, for whom there is no widely
acceptable general theory\cite{Mukhi1}. Our interest in the present
work is focussed on particle production at low-$p_T$ valued $Pb+Pb$
interaction at relatively lower SPS energies. The prime object of
this paper is to study the $p_T$-dependence of the invariant
cross-sections for main vaieties of hadron secondaries with the help
of one or two models which do not typically fall under the Standard
variety and which will be outlined in the next section. The main
species (like kaon, pion, proton) of the secondaries here are pions,
kaons and proton-antiproton produced in both lead-lead interactions
at $\sqrt{S_{NN}}$=17.3GeV and $E_{lab}$=20A, 30A, 40AGeV and Au+Au
reactions at RHIC at $\sqrt{S_{NN}}$=19.6GeV.
\par In our approach we would try (i) to demolish the artifact of
soft-hard division and try to apply a unified outlook by treating
the production of particles on a uniform footing.(ii) to check
whether the outlook of the divide between the exponentialisation and
the power-law nature based on the artificial soft-hard boundary is
of any real merit or worth and in probing this point we have been
spurred on to by a very recent report published by
Busza\cite{Busza1} with emphasis on the first lesson to be learnt
from PHOBOS Collaboration that there is no anomalous production of
low-$p_T$ (Soft) particles at RHIC; (iii) to study how the nature of
$p_T$-scaling in the particle production scenario manifests itself
in the extreme case of low-$p_T$ and the relatively lower side of
the high energy domain i.e., at SPS region.
\par The work to be reported here aims at putting a strong
question-mark to the time honoured contention of the large bulk of
the high energy physicists that the data on `soft' collisions could
be described by the exponential models; the power laws are
applicable for explaining the data on `hard' collisions. In fact, in
this work we have contested this view and have attempted at showing
that power laws could be applied almost universally.
\par The organisation of the work is as follows. In Section-2 we
present a brief outline of the very simple theoretical frameworks.
The Section-3 embodies the results of the work in graphical plots,
and tables which show up the used values of the parameters. In
Section-4 we have discussed in detail the very minute points that
worked in arriving at the desired results and had provided answers
to some anticipatory and probable questions from the readers. The
last section offers precisely the summary and outlook.
\section{Outlines of the Theoretical Framework } This
section is divided into subsections comprising the (i)
Power-law-based $p_T$-scaling Model (also named Hagedorn Model) and
(ii) The Combinational Approach, both of which are very much in use
in recent times by some of us
\subsection{Hagedorn Model : Essentially a Power Law with $p_T$-scaling}
Our objective here is to study the inclusive $p_T$-spectra of the
various secondaries of main varieties produced in $pp$ collisions.
The kinematics of an inclusive reaction
$h_ah_b$$\longrightarrow$$hX$ is described by Lorentz invariants.
These are e.g. the center-of-mass energy squared s =
($P_a$+$P_b$)$^2$, the transverse momentum transverse (squared) t =
($P_h$-$P_a$)$^2$ and the missing mass $M_X$, where $M_X$ is the
mass of the undetected components that are produced in the reaction
process, denoted by `X' in $h_ah_b$$\longrightarrow$$hX$. It is
common to introduce the dimensionless variables (u = $M_X$$^2$-s-t),
\begin{equation}
     x_1 = -\frac {u}{s},
     x_2 = -\frac {t}{s}
\end{equation}
where s,t,u are called Mandelstam variables. These variables are
related to the rapidity y and radial scaling factor, $x_R$ of the
observed hadron by
\begin{equation}
     y = \frac{1}{2}log(x_1/x_2)
\end{equation}
\par
\begin{equation}
    x_R = \frac {2|\overrightarrow{p_{cm}}|}{\sqrt{s}} = 1 - \frac {M_X^2}{s} = x_1+x_2
\end{equation}
Since most of existing data are at y = 0 where $x_R$ = $x_T$ =
2$p_T$/$\sqrt{s}$ and $|\overrightarrow{p_{cm}}|$ is just the
magnitude of the three momentum in c.m. system, one often refers to
the scaling of the invariant cross section as "$x_T$ scaling". For
y$\neq$0, we find the variable $x_R$ more useful than $x_T$, since
$x_R$ allows a smooth matching of inclusive and exclusive reactions
in the limit $x_R$ $\longrightarrow$ 1.
 \par We will assume that at high $p_T$, the inclusive cross section takes a factorized form.
 And one such a factorized form was given by Back et al\cite{Back1}.
\begin{equation}
     \frac {dN}{dp_T}= \frac {p_T(n-2)(n-1)}{{p_0}^2}(1+\frac{p_T}{p_0})^{-n}
\end{equation}
where n and $p_0$ are adjustable parameters. The values of the
exponent n are just numbers. $p_0$ is to be viewed as a energy-band
dependent critical value of the transverse momentum within the
low-$p_T$ limits and is introduced for the sake of making the term
within the parenthesis dimensionless, thus lending the expression
within the parenthesis a scaling form with $p_T/p_0$, called
$p_T$-scaling.
 \par With the simplest recasting of form the above expression (4) and with replacements
 like $p_T$=x, $p_0$=q, C = a normalisation factor, and in the light of the definitions of
 the inclusive cross-sections, we get the following form as the final working formula
\begin{equation}
 z=f(x)=\frac{C}{q^2}(1+\frac{x}{q})^{-n}
\end{equation}

\subsection{Combinational Approach : De-Bhattacharyya Model}
\par
The combinational approach outlines a method for arriving at the
results to be obtained on some observables measured in
particle-nucleus or nucleus-nucleus interactions at high energies
from those obtained for the basic nucleon-nucleon (or proton-proton)
collision. And the results for nucleon-nucleon interactions are
based on power-law fits which are assumed to be physically
understood somewhat fairly in the light of both thermal model and/or
pQCD-related phenomenology. So, essentially this represents a
notional combination of power law model for pp reactions and the the
introduction of some mass number-dependent product term signifying
the nuclear effects on invariant cross-sections, from which the name
`Combinational Approach' is derived. And this notional combination
subsumes the property of factorization which is one of the cardinal
principles in the domain of particle physics.
\par
The expression for the transverse momentum-dependence of the
inclusive cross-section for secondary particle, Q, produced in
nucleus-nucleus(AB) collisions is given by\cite{De1}-\cite{De2}
\begin{equation}
E\frac{d^3\sigma}{dp^3}|_{AB \rightarrow QX} \sim (AB)^{f(p_T)}E\frac{d^3\sigma}{dp^3}|_{pp \rightarrow QX}
\end{equation}
Where A and B on the right-hand side of the above equation stand for
the mass numbers of two colliding nuclei; the term,
$E\frac{d^3\sigma}{dp^3}|_{pp \rightarrow QX}$ is the inclusive
cross section for production of the same secondary,Q,in pp or
$p\overline{p}$ collision at the same (center-of-mass) c.m. energy.
\par The nature of ($p_T$)-dependence of the inclusive cross-section term,
$E\frac{d^3\sigma}{dp^3}|_{pp \rightarrow QX},$ occurring in
eqn.(1), for production of a Q-species in $pp/p\overline{p}$
reactions at high energies is taken here in the form of a power-law
as was initially suggested by G.Arnison et al.\cite{Arnison1}:
\begin{equation}
E\frac{d^3\sigma}{dp^3}|_{pp \rightarrow QX}\approx C_1 (1+\frac{p_T}{p_0})^{-n}
\end{equation}
Where $C_1$ is the normalization constant; and $p_0$ and n are two
interaction-dependent parameters for which the values are to be
obtained by fitting the pp and $p\overline{p}$ data at various
energies. Of course, such a power-law form was applied to understand
the nature of the transverse momentum spectra of the pion
secondaries by some other authors\cite{Bocqet1}-\cite{Peitzmann1} as
well.
\par Hence including eqn. (7) and a parametrization for the
factor, $f(p_T)$, into eqn. (6), the final working formula is given
by\cite{De1}-\cite{De2}
\begin{equation}
E\frac{d^3\sigma}{dp^3}|_{AB \rightarrow QX} \approx C(AB)^{(\alpha p_T - \beta p^2_T)} (1+\frac{p_T}{p_0})^{-n}
\end{equation}
Where C,$\alpha$ and $\beta$ are constants and have to be determined
by fitting the measured data on ($p_T$)-spectra for production of
charged hadrons in nucleus-nucleus collisions at high energies. Some
sort of physical interpretations for $\alpha$ and $\beta$ are given
in some of our previous works\cite{De3},\cite{De4}.
\par A useful way\cite{Back1},\cite{Arsene1} to compare the spectra from
nucleus-nucleus collisions to those from nucleon-nucleon collisions
is to scale the normalized pp (or $p\overline{p}$)spectrum (assuming
the value of inelastic pp cross-section, $\sigma^{pp}_{inel}
\approx$ 41 mb) by the number of binary collisions,$\langle
N_{coll}\rangle$, corresponding to the centrality cuts applied to
the nucleus-nucleus spectra and construct the ratio. This ratio is
called the nuclear modification factor, $R_{AB}$, which is to be
expressed in the form
\begin{equation}
R_{AB} =
\frac{\frac{1}{\langle N_{coll} \rangle}E\frac{d^3N}{dp^3}|_{AB}}{\frac{1}{\sigma^{PP}_{inel}}E\frac{d^3\sigma}{dp^3}|_{PP}}
\end{equation}
\par It is to be noted here that both the numerator and the
denominator of equation (9) contain a term of the form
$(1+\frac{p_T}{p_0})^{-n}$which gives the $p_T$-dependence of the
hadronic-spectra produced in basic ($pp/p\overline{p}$)collision.
And as the other terms like, $\langle N_{coll}\rangle,
\sigma^{pp}_{inel}$ are constants for a specific interaction at a
definite energy and fixed centrality, we can obtain by combining
eqn.(8) and (9) the final expression for the ratio value in the
following form :
\begin{equation}
R_{AB} \propto (AB)^{(\alpha p_T - \beta p^2_T)}
\end{equation}
\par The above steps provide the operational aspects of the
combinational approach (CA). But, this approach outlines a method
for arriving at the result to be obtained on some observables
measured in particle-nucleus or nucleus-nucleus interactions at high
energies from those obtained exclusively for only the basic
nucleon-nucleon (or proton-proton)collision.And results for
nucleon-nucleon collisions are based on power-law form (eqn.(7))
which is supposed here to be fairly physically understood in the
light of both thermal model and/or pQCD-related phenomenology.
Besides, if the data sets on a specific observable are measured in
pp reactions at five or six different high energies at reasonably
distant intervals, the pure parameter-effects on $p_0$ and n could
be considerably reduced and we may build up a methodology for
arriving at $p_0$ and n values at any other different energy by
drawing the graphical plots of $p_0$ versus $\sqrt{S}$ and n versus
$\sqrt{S}$ curves, as are done in some of our previous
works\cite{De1},\cite{De3}. And on this supposition of availability
of data in PP reactions at certain intervals of energy values, the
number of arbitrary parameters for NA or AA collisions is reduced to
only three which offer us quite a handy, useful and economical tool
to understand the various aspects of the data characteristics.
\par The $p_0$ and n values in eqn. (7) represent the contributions from
basic NN (PP OR $p\overline{p}$) collision at a particular energy.
The values of $p_0$ and n are to be obtained from the expressions
and plots shown in the work of De et al\cite{De1},\cite{De3}. The
relevant expressions are
\begin{equation}
p_0(\sqrt{S_{NN}}) = a + \frac{b}{\sqrt{\frac{S_{NN}}{GeV^2}}\ln(\sqrt{\frac{S_{NN}}{GeV^2}})}
\end{equation}
\begin{equation}
n(\sqrt{S_{NN}}) = a' + \frac{b'}{\ln^2(\sqrt{\frac{S_{NN}}{GeV^2}})}
\end{equation}
\par The values of the parameters $a$, $a'$, b and $b'$ for different
secondaries are taken from Ref.\cite{De2}. The empirically proposed
nature of the plots based on eqn. (11) and eqn. (12) against the
data-sets, have also been presented for $\pi^\pm$, $K^\pm$ and
$p^\pm$ production separately in a subsequent section (Section 4).
\section{Results}
In order to attain the comparability of the results obtained at
various energies, we have, at the very beginning, converted all the
relevant energies (with laboratory energies) into the c.m. system.
And they have been presented in a tabular form as is given in Table
1. The results are presented here in graphical plots and the
accompanying tables for the values of the used parameters. In
Fig.(1a) and Fig.1(b) the differential cross-sections for negative
and positive pion, kaon and proton-antiproton production cases in
Pb+Pb collisions at SPS energies are reproduced by the used
empirical parametrizations given in expressions (4) and (7). The
parameter values used to obtain the model-based plots are shown in
Table-2. The figures in all the cases have been appropriately
labeled and the parameters are shown in the tables as mentioned in
the text for each case. The data are obtained at such low-$p_T$
values and at the relatively lower side of the high energy in terms
of $m_T$-values, instead of $p_T$ values. The relationship between
$m_T$ and $p_T$ is generally given by $m_T^2$=$m_h^2+p_T^2$. Within
the low-$p_T$ limits and for the low-mass particles ($m_h\ll p_T$)
produced in any high energy collisions $m_T$$\approx$$p_T$. As the
measured data are obtained and exhibited with $m_T$ in the abscissa
we choose to retain them intact; the ordinate-observable too is kept
thus undisturbed. The plots in Fig.2(a) and Fig.2(b) are for
positive and negative pion, positive and negative kaon in Pb+Pb
collisions at 30A GeV and the corresponding parameters are depicted
in Table-3. The plots presented in Fig.(2c) for proton-antiproton
production in lead-lead interaction at SPS energies, specifically at
17.3 GeV and the corresponding parameters are depicted in Table-3.
The graphs in Fig.3 and Fig.4 present the results for the
secondaries $\pi^\pm$, $K^\pm$ and proton-antiproton produced in the
Au+Au collisions at 19.6 GeV. Used parameter values
 for these two figures are given in Table-4. The rest of the figures
 demonstrated in the Fig.5, Fig.6, Fig.8 and Fig.9 are for collisions
as labelled and are based on the De-Bhattacharyya model for the
major varieties of hadronic secondaries produced at four very close
energies as mentioned in each of the figures separately. The
parameter values used to obtain the nature of fits are shown in
Table-5,6,9. The graph plotted in Fig.7 are based on both Power Law
Model and DBP Model respectively, at 40A GeV in Pb+Pb collsions. The
parameter values obviously remain the same as given in table-7 and
Table-8.

\par The graphs plotted in Fig. 10(a) and Fig. 11(a) are based on
Power Law Model and DBP Model respectively; they represent the fits
to the invariant $p_T$-spectra for production of the main varieties
of secondaries in 19.6 GeV Au+Au collisions at mid-rapidity and at
the range of 0-10\% centrality which covers the highest centrality
region of nuclear collisions. The parameter values used are given in
Table-10 and Table-11. The rest of the plots in Fig. 10 and Fig. 11
are for the nature of the charge-ratios-behaviours for the specific
variety of the secondary particles. The Fig. 10(b), Fig. 10(c) and
Fig. 10(d) are drawn on the basis of the Power Law Model; and the
plots shown in Fig. 11(b), Fig. 11(c) and Fig. 11(d) are on the
basis of the DBP model for the same set of charge-ratios.
\par
At last, even on the basis of a very few four close-ranging energies
and on an approximation that the measured data-values on Pb+Pb and
Au+Au at the neighboring would not differ too much, we have tried to
check here the merits of the phenomenological energy-dependences
proposed by us in eqns. (11), (12) of the two key parameters, viz,
$p_0$ and n. In Fig. 12 to Fig. 14 we have plotted the parameter
values $p_0$ along the Y-axis with the energy-values as the
corresponding X-axis. The dotted curves in all these figures depict
the nature of the parameters obtained by the proposed empirical
expressions represented by eqns. (11) and (12). The
phenomenologically formulated expressions reproduce fairly well the
nature of fit-values used for the two parameters. Despite the
limitations pointed here out before and some others, the agreements
are modestly encouraging.
\section{General Discussion and Some Specific Points}
By all indications, the results manifested in the measured data on
the specific observables chosen here are broadly consistent with
both the approaches put into work here. This is modestly true of
even the nature of charge-ratios which provide virtually a
cross-check of the models utilised here. Of course, at this point
let us make some comments on our model-based plots, especially the
plots on the charge-ratios . This is not very surprising in the
sense that the De-Bhattacharyya phenomenology (DBP) through a
parametrization is also essentially a power law model with a bare
A-dependence, while in the pure power law model any of the chosen
parameters absorb the nuclear-dependence. Still, the DBP-model
inducts some physical postulates which are as follows :(i) It is
assumed that the inclusive cross section of any particle in a
nucleus-nucleus (AB) collision can be obtained from the production
of the same in nucleon-nucleon collisions by multiplying by a
product of the atomic numbers of each of the colliding nuclei raised
to a particular function, $\epsilon(y,p_T)$, which at first is
unspecified (Equation 6), (ii) Secondly, we have accepted that
factorization \cite{Sau1} of the function $\epsilon(y,p_T)$ =
$f(y)g(p_T)$ which helps us to perform the integral over $p_T$ in a
relatively simpler manner, (iii) Finally, we have based on the
ansatz that the function f(y) can be modeled by a quadratic function
with the parameters $\alpha$ and $\beta$ (Equation 10). They are to
be tested in the high energy experiments of the future generations.
Besides, the present DBP-based approach
 to deal with the data advances a systematic methodical
approach in which the main parameters could be determined and
ascertained, if and only if, measured laboratory data of higher
accuracy and precision are available at different energies in a
regular and successive interval. So, the lack of predictivity of the
used two models is caused only by the circumstances, i.e. the lack
of measured data at the successive and needed intervals ; the
problem can be remedied by supplying the necessary and reliable data
from the arranged laboratory experiments at high - to - very high
energies. However, the problem of constraining the parameters still
remains. The other observations are : as is expected for two
symmetric collisions of neighbouring values of mass numbers at very
close energies, the measured data do not reveal any significant
differences with respect to the observables chosen by the
experimental groups. This work demonstrates somewhat convincingly
that the power law models can easily take care of data even for very
low-$p_T$ (soft) collisions; so the notion of compartmentalisation
between the possible applicability of the power law models and of
the exponential models is only superficial. Besides, the power law
models which establish them as more general ones obtain a clear edge
over the exponential models. Reliable data on various other related
observables are necessary for definitive final conclusions. By all
indications, the experimentally observed nature of $p_T$- scaling is
found to remain valid even in the studied low-$p_T$ range of this
paper. The deeper physical implications of this work have implicity
been pointed out in the last two paragraphs of the Section 1 of this
work, wherein we made very clear and categorical statements on our
intentions, purposes and the prime objectives. So, in order to avoid
repetitions, we refrain ourselves here very carefully from commiting
a rehash of the same. However, let us now try to pinpoint below some
physical points and considerations which provide the necessary
underpinning of the power law models.
\par
The wide and near successful applications of the various forms of
power laws have, by now, grown almost universal in almost all the
branches of physics as well as in other branches of sciences.
Speaking in the most general and scientific terms, the processes
which are complex, violent and dissipative contributing to the
non-equilibrium phenomena do generally subscribe to the power laws.
The used power law behaviors are commonly believed to be the
``manifestations of the dynamics of complex systems whose striking
feature is of showing universal laws characterized by exponents in
scale invariant distributions that happen to be basically
independent of the details in the microscopic
dynamics\cite{Lehmann1}". Now let us revert from the general to the
particular case(s) of high energy particle-particle,
particle-nucleus or nucleus collisions. For purely hadronic,
hadronuclear or nuclear interactions one of the basic features is :
irrespective of the initial state, agitations caused by the
impinging projectile (be it a parton or a particle/nucleon) generate
system effects of producing avalanches of new kind of partons
[called quark(s)/gluon(s)/any other(s)] which form an open
dissipative system. The avalanches caused by production of excessive
number of new partons give rise to the well-known phenomenon of
jettiness of particle production processes and of cascadisation of
the particle production processes leading to the fractality as was
shown in a paper by Sarcevic\cite{Sarcevic1}. These cascades are
self-organizing, self-similar and do just have the fractal behavior.
Driven by the physical impacts of these well-established factors, in
the high energy collision processes do crop up the several
power-laws. And how such power laws do evolve from exponential
origins or roots is taken care of by the induction of Tsallis
entropy\cite{Biro1} and a generalisation of Gibbs-Boltzmann
statistics for long-range and multifractal processes. Following
Sarcevic\cite{Sarcevic1} the relationship between/among
cascadisation, self-similarity and fractality is/was evinced in a
paper\cite{Biswas1} by a set of the present authors. In this paper
we have, once more, tried to examine the worth and utility of such
power laws as have been advocated here.
\par
In the data-plots of Fig. 3(a) we observe some differences between
the trends of WA98 data and the others. But it is to be noted that
the data in WA98 experiment is for the observation of neutral pions,
whereas our plot is intended to be one of the charged varieties of
pions, e.g., $\pi^+$ detected at the experimental energy 17.3 GeV in
c.m. system; besides, the STAR data was measured at 19.6 GeV (c.m.).
This apart, in the WA98 experiment the observable along the Y-axis
was just $E\frac{d^3\sigma}{dp^3}$, whereas for the others the
Y-axis was described to be $\frac{1}{2\pi}\frac{d^2n}{m_T dm_Tdy}$.
The data from WA98 took up values of $m_T-m_0$ from 1 to 4 GeV/c.
But all the other plots were limited to just $m_T-m_0$=1 GeV/c. So
there are a host of factors of differences between WA98 data and the
rest. The differences in the magnitudes of the $\pi^+$ plots in the
invariant cross section between various data-sets might be a
cumulative effect of all the above-mentioned factors.
\par
Quite spectacularly, in fig. 3(d) for production of $K^-$, the
differences between the data-points measured by WA98 and the rest
are surely non-negligible. So the existence of discrepancy to a
certain degree cannot be denied altogether.
\par
However, one must note that the observable plotted as ordinate in
the WA98 experiment is a bit different from the others, where does
occur a term, denoted by $N_{evt}$ related with both statistical and
systematic uncertainties, though the uncertainties in all other
cases (i.e., for non-WA98) the errors are only statistical. Our
guess is : the twin factors of systematic uncertainties and the
separate y-observable [in the WA98] along the ordinate are
responsible for introducing such differences as are shown in
data-plot of Fig. 3(d). The reason(s) might be something else as
well.
\par
We observe that the data points on invariant cross-sections for
production of protons show relatively much slower fall with $m_T$
($p_T$) than the other prime varieties of hadrons. Thus one might
have doubts on the accuracy of data-measurements and recordings for
protons. If this is too unlikely to be the case, we guess that in
the detection/measurement process the part of the `leading' protons
has disguised themselves and appeared as the product
proton-particles enhancing protonic invariant cross-section for
which the fall in the invariant cross-section with $m_T$ is much
less. Uptil now, this explanation is certainly just tentative.
\par In the end, one more comment is in order. Quite knowingly,
we have used here the usual binned $\chi^2$-method with the
attending limitations of this approach to check the goodness-of-fit
of our results to the data, as unbinned multivariate goodness-of-fit
tests\cite{Williams1} have not yet gained much ground in the High
Energy Physics (HEP) sector.
\section{Concluding Remarks}
Let us we present here very precisely the main conclusions of this
work. Firstly, for these limited sets of data on production of soft
particles in high energy collisions which have, so far, defied
explanation, we have attempted to provide two alternative
theoretical/phenomenological approaches for their interpretation in
modestly successful manners. Secondly, in fact, these two approaches
conceptually and inwardly are somewhat interlinked, for which
limited successes of both of them are not very surprising. Thirdly,
power law models are seen to act much better here than all other
models; besides, they are much more general than the others.
Fourthly, the applications of power law models are quite widespread
in the different fields of physics in particular and of science in
general, for which active interests in investigating the origin of
these power laws have been aroused. And this has, so far, given rise
to, in the main, two parallel streams of thought, of which one is
the cascadisation phenomena and fractal mechanisms; and the other is
the science of nonequilibrium phenomena that are generally probed by
applying the Tsallis entropy and Tsallis statistics\cite{Tsallis1}.
Confirmation of such multiple educated guesses can be made only by
further dedicated researches in these fields.
\par At last, in response to what we learn very precisely from this work we
submit the following few points: (i) Our unquestioned belief in and
reliance on the Standard Model(SM) have, so far, been virtually
'regimented', for which we fail to think of any other avenues and
accept the singularity and uniqueness phenomena of the SM as taken
for granted. (ii) In the sphere of surely very limited sets of data
we explore and assess here the potential of two alternative models
in explaining the observed data. (iii) And as we have succeeded in
our attempts to a considerable extent, in our opinion, these two
models dealt herewith could in future be viewed and projected as
possible alternative approaches to explain the nature of observed
and measured data-sets on `soft' production of particle in high
energy collisions.

\begin{center}
\par{\textbf{Acknowledgements}}
\end{center}
\par
The authors express their deep indebtedness to the learned referees
for some encouraging remarks, constructive comments and valuable
suggestions which helped a lot in improving an earlier version of
the manuscript.

\singlespacing

\newpage
{\singlespacing{
\begin{table}
\begin{center}
\begin{small}
\caption{Conversion of energy system}
\begin{tabular}{|c|c|c|c|}\hline
Beam Energy & $20A GeV$& $30A GeV$  & $40A GeV$ \\
\hline
$\sqrt{S_{NN}}$&$6.3GeV$&$7.6GeV$&$8.7GeV$\\
\hline
\end{tabular}
\end{small}
\end{center}

\begin{center}
\begin{small}
\caption{ Numerical values of the fit parameters of power law
equation for $Pb+Pb$ Collisions.}
\begin{tabular}{|c|c|c|c|c|c|}\hline
Beam Energy & $Products$& $c$  & $q(GeV/c)$ & $n$ & $\frac{\chi^2}{ndf}$\\
\hline
$20A GeV$&$\pi^-$&$43.935\pm0.015$&$1.197\pm0.037$&$10.010\pm0.234$&$0.612/7$\\
\hline
$20A GeV$&$\pi^+$&$38.950\pm0.233$&$1.178\pm0.087$&$9.410\pm0.531$&$2.628/6$\\
 \hline
$20A GeV$&$K^-$&$1.690\pm0.013$&$2.404\pm0.014$&$14.003\pm0.054$&$0.806/4$\\
\hline
$20A GeV$&$K^+$&$5.515\pm0.136$&$2.001\pm0.022$&$11.888\pm0.180$&$4.894/4$\\
\hline
$20A GeV$&$p$&$3.568\pm0.062$&$2.757\pm0.138$&$8.006\pm0.352$&$0.904/4$\\
\hline
\end{tabular}
\end{small}
\end{center}

\begin{center}
\begin{small}
\caption{ Numerical values of the fit parameters of power law
equation for $Pb+Pb$ Collisions.}
\begin{tabular}{|c|c|c|c|c|c|}\hline
Beam Energy & $Products$& $c$  & $q(GeV/c)$ & $n$ & $\frac{\chi^2}{ndf}$\\
\hline
$30A GeV$&$\pi^-$&$52.202\pm0.307$&$1.158\pm0.073$&$9.232\pm0.438$&$1.158/5$\\
\hline
$30A GeV$&$\pi^+$&$47.041\pm0.637$&$1.344\pm0.211$&$10.026\pm1.166$&$5.790/8$\\
 \hline
$30A GeV$&$K^-$&$2.551\pm0.040$&$2.001\pm0.021$&$11.240\pm0.115$&$3.519/3$\\
\hline
$30A GeV$&$K^+$&$8.962\pm0.341$&$1.999\pm0.017$&$12.986\pm0.293$&$3.697/4$\\
\hline
$17.3 GeV$&$p$&$15027.3\pm34.99$&$2.047\pm0.0001$&$8.504\pm0.011$&$1.252/4$\\
\hline
$17.3 GeV$&$\overline{p}$&$0.450\pm0.042$&$0.439\pm0.049$&$3.999\pm0.037$&$1.338/5$\\
\hline
\end{tabular}
\end{small}
\end{center}

\begin{center}
\begin{small}
\caption{ Numerical values of the fit parameters for pion, kaon,
proton and antiproton using Power Law Model for $Au+Au$ collisions
at 19.6GeV.}
\begin{tabular}{|c|c|c|c|c|c|}\hline
Beam Energy & $Products$ & $c$  & $q(GeV/c)$ & $n$ & $\frac{\chi^2}{ndf}$\\
\hline
$19.6GeV$&$\pi^+$&$12.588\pm0.183$&$0.559\pm0.061$&$5.547\pm0.358$&$6.518/7$\\
\hline
$19.6GeV$&$\pi^-$&$13.443\pm0.089$&$2.001\pm0.011$&$13.192\pm0.065$&$27.759/29$\\
 \hline
$19.6GeV$&$K^+$&$1.455 \pm0.017 $&$2.002 \pm0.256 $&$9.803 \pm0.109 $&$13.526/11 $\\
\hline
$19.6GeV$&$K^-$&$0.742\pm0.007$&$6.640\pm0.038$&$29.998\pm0.050$&$27.027/18$\\
\hline
$19.6GeV$&$p$&$1.995\pm0.027$&$0.887\pm0.103$&$4.257\pm0.233$&$18.933/11$\\
\hline
$19.6GeV$&$\overline{p}$&$0.188\pm0.005$&$2.440\pm0.057$&$7.001\pm0.044$&$2.741/06$\\
\hline
\end{tabular}
\end{small}
\end{center}
\end{table}

\begin{table}
\begin{center}
\begin{small}
\caption{Numerical values of the fit parameters of DBP equation for
$Pb+Pb$ Collisions.}
\begin{tabular}{|c|c|c|c|c|c|}\hline
Beam Energy & $Products$& $c$  & $\alpha$ & $\beta$ & $\frac{\chi^2}{ndf}$\\
\hline
$20A GeV$&$\pi^-$&$1608.85\pm18.320$&$0.020\pm0.008$&$0.004\pm0.010$&$20.330/12$\\
\hline
$20A GeV$&$\pi^+$&$1492.27\pm61.830$&$0.031\pm0.017$&$0.018\pm0.018$&$19.790/10$\\
 \hline
$20A GeV$&$K^-$&$35.667\pm0.151$&$0.202\pm0.010$&$0.083\pm0.010$&$9.847/4$\\
\hline
$20A GeV$&$K^+$&$94.412\pm2.063$&$0.334\pm0.013$&$0.211\pm0.014$&$2.456/3$\\
\hline
$20A GeV$&$p$&$11.309\pm4.938$&$0.522\pm0.106$&$0.142\pm0.066$&$2.148/4$\\
\hline
\end{tabular}
\end{small}
\end{center}

\begin{center}
\begin{small}
\caption{ Numerical values of the fit parameters of DBP equation for
$Pb+Pb$ Collisions.}
\begin{tabular}{|c|c|c|c|c|c|}\hline
Beam Energy & $Products$& $c$  & $\alpha$ & $\beta$ & $\frac{\chi^2}{ndf}$\\
\hline
$30A GeV$&$\pi^-$&$1707.12\pm18.260$&$0.059\pm0.007$&$0.029\pm0.009$&$21.159/12$\\
\hline
$30A GeV$&$\pi^+$&$1697.79\pm27.510$&$0.050\pm0.002$&$0.026\pm0.004$&$24.543/15$\\
 \hline
$30A GeV$&$K^-$&$45.951\pm1.407$&$0.302\pm0.013$&$0.181\pm0.013$&$15.312/12$\\
\hline
$30A GeV$&$K^+$&$104.667\pm2.330$&$0.373\pm0.011$&$0.244\pm0.012$&$5.078/5$\\
\hline
$17.3 GeV$&$p$&$10.352\pm0.070$&$0.654\pm0.002$&$0.121\pm0.001$&$0.635/8$\\
\hline
$17.3 GeV$&$\overline{p}$&$6.084\pm0.121$&$0.723\pm0.009$&$0.281\pm0.009$&$2.259/8$\\
\hline
\end{tabular}
\end{small}
\end{center}

\begin{center}
\begin{small}
\caption{ Numerical values of the fit parameters for negative pion,
positive and negative kaon, proton and antiproton using DBP Model
for $Pb+Pb$ collisions at 40A GeV.}
\begin{tabular}{|c|c|c|c|c|c|}\hline
Beam Energy & $Products$ & $c$  & $\alpha$ & $\beta$ & $\frac{\chi^2}{ndf}$\\
\hline
$40A GeV$&$\pi^-$&$1302.44\pm9.467$&$0.079\pm0.008$&$0.065\pm0.012$&$0.109/15$\\
\hline
$40A GeV$&$K^+$&$108.27\pm0.428$&$0.220\pm0.002$&$0.090\pm0.003$&$0.328/13$\\
\hline
$40A GeV$&$K^-$&$44.963\pm0.323$&$0.200\pm0.005$&$0.093\pm0.006$&$0.099/13$\\
\hline
$40A GeV$&$p$&$80.879\pm1.260$&$0.346\pm0.007$&$0.109\pm0.006$&$0.638/14$\\
\hline
$40A GeV$&$\overline{p}$&$0.599\pm0.004$&$0.382\pm0.005$&$0.145\pm0.006$&$0.129/14$\\
\hline
\end{tabular}
\end{small}
\end{center}

\begin{center}
\begin{small}
\caption{Numerical values of the fit parameters for negative pion,
positive and negative kaon, proton and antiproton using Power Law
Model for $Pb+Pb$ collisions at 40A GeV.}
\begin{tabular}{|c|c|c|c|c|c|}\hline
Beam Energy & $Products$ & $c$  & $q(GeV/c)$ & $n$ & $\frac{\chi^2}{ndf}$\\
\hline
$40A GeV$&$\pi^-$&$54.328\pm1.296$&$1.198\pm0.022$&$10.003\pm0.042$&$0.103/15$\\
 \hline
$40A GeV$&$K^+$&$9.176\pm0.166 $&$1.999 \pm0.007 $&$13.856 \pm0.132 $&$0.617/07 $\\
\hline
$40A GeV$&$K^-$&$2.429\pm0.067$&$1.734\pm0.048$&$10.031\pm0.210$&$1.114/11$\\
\hline
$40A GeV$&$p$&$12.827\pm0.589$&$1.629\pm0.043$&$10.002\pm0.070$&$0.433/08$\\
\hline
$40A GeV$&$\overline{p}$&$0.078\pm0.002$&$1.382\pm0.054$&$7.022\pm0.178$&$0.800/08$\\
\hline
\end{tabular}
\end{small}
\end{center}
\end{table}

\begin{table}
\begin{center}
\begin{small}
\caption{ Numerical values of the fit parameters for pion, kaon,
proton and antiproton using DBP Model for $Au+Au$ collisions at
19.6GeV.}
\begin{tabular}{|c|c|c|c|c|c|}\hline
Beam Energy &$Products$ & $c$  & $\alpha$ & $\beta$ & $\frac{\chi^2}{ndf}$\\
\hline
$19.6GeV$&$\pi^+$&$223.493\pm28.050$&$0.239\pm0.037$&$0.100\pm0.027$&$4.202/6$\\
\hline
$19.6GeV$&$\pi^-$&$352.948\pm12.490$&$0.113\pm0.006$&$0.029\pm0.002$&$3.292/5$\\
\hline
$19.6GeV$&$K^+$&$22.951\pm0.709$&$0.350\pm0.014$&$0.168\pm0.015$&$9.672/8$\\
\hline
$19.6GeV$&$K^-$&$13.611\pm0.041$&$0.287\pm0.005$&$0.110\pm0.006$&$19.681/19$\\
\hline
$19.6GeV$&$p$&$7.030\pm0.044$&$0.648\pm0.004$&$0.168\pm0.006$&$15.994/14$\\
\hline
$19.6GeV$&$\overline{p}$&$0.557\pm0.021$&$0.649\pm0.024$&$0.297\pm0.032$&$4.117/4$\\
\hline
\end{tabular}
\end{small}
\end{center}

\begin{center}
\begin{small}
\caption{ Numerical values of the fit parameters for pion, kaon,
proton and antiproton using Power Law Model for $Au+Au$ collisions
at 19.6GeV for 0-10\% centrality.}
\begin{tabular}{|c|c|c|c|c|c|}\hline
Beam Energy&$Products$ & $c$  & $q(GeV/c)$ & $n$ & $\frac{\chi^2}{ndf}$\\
\hline
$19.6GeV$&$\pi^+$&$18.522\pm0.520$&$0.592\pm0.019$&$6.689\pm0.156$&$20.538/13$\\
\hline
$19.6GeV$&$\pi^-$&$15.410\pm0.302$&$0.658\pm0.026$&$6.834\pm0.216$&$13.235/13$\\
 \hline
$19.6GeV$&$K^+$&$1.686\pm0.028 $&$1.833 \pm0.070 $&$8.648 \pm0.234 $&$0.841/19 $\\
\hline
$19.6GeV$&$K^-$&$0.999\pm0.039$&$1.999\pm0.032$&$9.123\pm0.264$&$0.700/12$\\
\hline
$19.6GeV$&$p$&$1.672\pm0.016$&$1.567\pm0.025$&$6.587\pm0.066$&$10.212/16$\\
\hline
$19.6GeV$&$\overline{p}$&$0.159\pm0.002$&$1.568\pm0.018$&$6.587\pm0.052$&$8.980/21$\\
\hline
\end{tabular}
\end{small}
\end{center}

\begin{center}
\begin{small}
\caption{ Numerical values of the fit parameters for pion, kaon,
proton and antiproton using DBP Model for $Au+Au$ collisions at
19.6GeV for 0-10\% centrality.}
\begin{tabular}{|c|c|c|c|c|c|}\hline
Beam Energy&$Products$ & $c$  & $\alpha$ & $\beta$ & $\frac{\chi^2}{ndf}$\\
\hline
$19.6GeV$&$\pi^+$&$435.425\pm20.210$&$0.353\pm0.021$&$0.085\pm0.022$&$11.887/26$\\
\hline
$19.6GeV$&$\pi^-$&$465.221\pm16.760$&$0.301\pm0.016$&$0.043\pm0.016$&$6.918/29$\\
 \hline
$19.6GeV$&$K^+$&$24.176\pm0.328 $&$0.484\pm0.004 $&$0.009\pm0.001 $&$10.611/19 $\\
\hline
$19.6GeV$&$K^-$&$14.898\pm0.274$&$0.469\pm0.005$&$0.018\pm0.012$&$8.199/13$\\
\hline
$19.6GeV$&$p$&$11.273\pm0.103$&$0.731\pm0.002$&$0.271\pm0.003$&$14.043/21$\\
\hline
$19.6GeV$&$\overline{p}$&$1.139\pm0.018$&$0.730\pm0.006$&$0.269\pm0.015$&$12.847/21$\\
\hline
\end{tabular}
\end{small}
\end{center}
\end{table}

\newpage
\begin{figure}
\subfigure[]{
\begin{minipage}{.5\textwidth}
\centering
\includegraphics[width=2.5in]{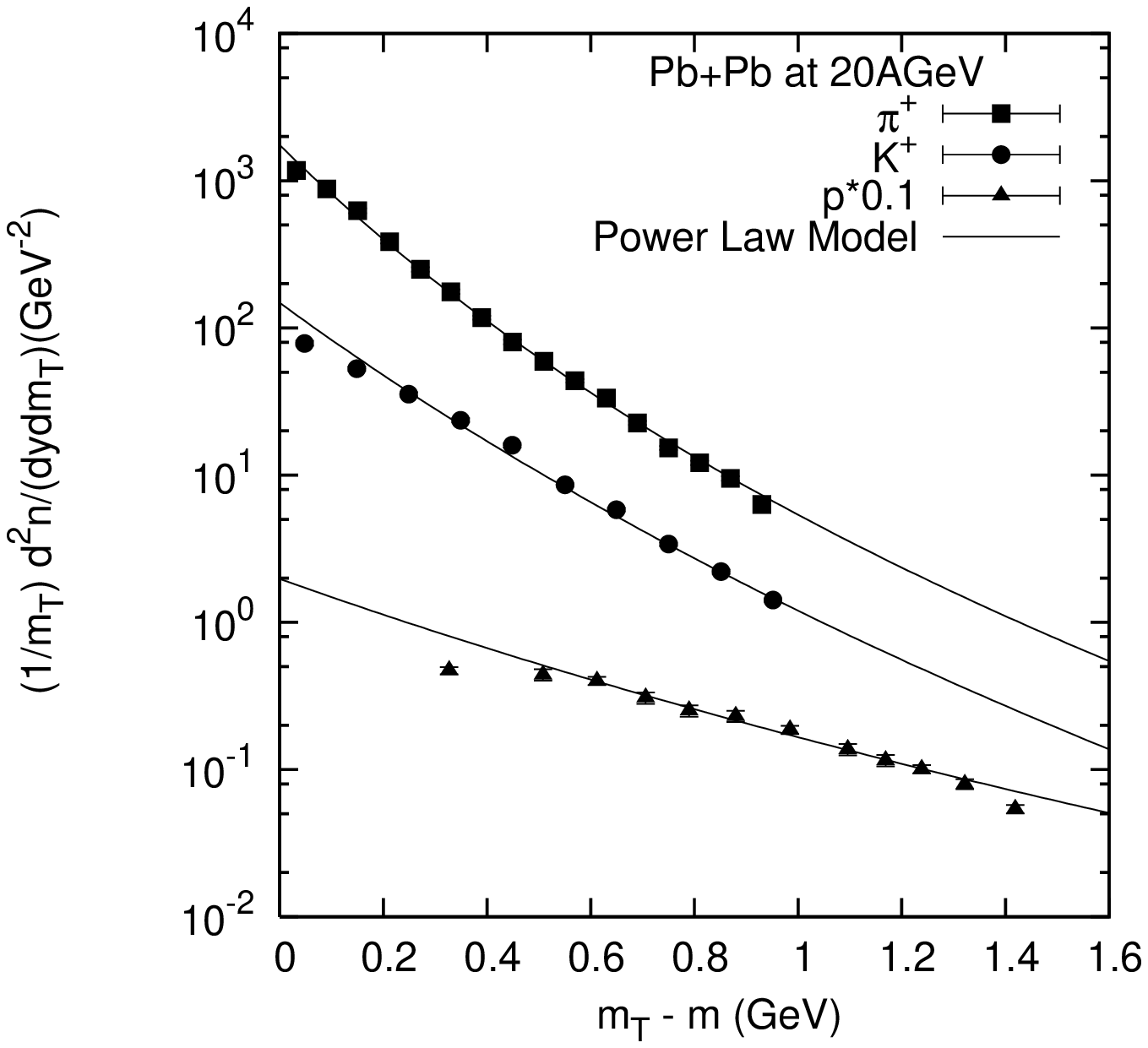}
\end{minipage}}%
\subfigure[]{
\begin{minipage}{.5\textwidth}
\centering
 \includegraphics[width=2.5in]{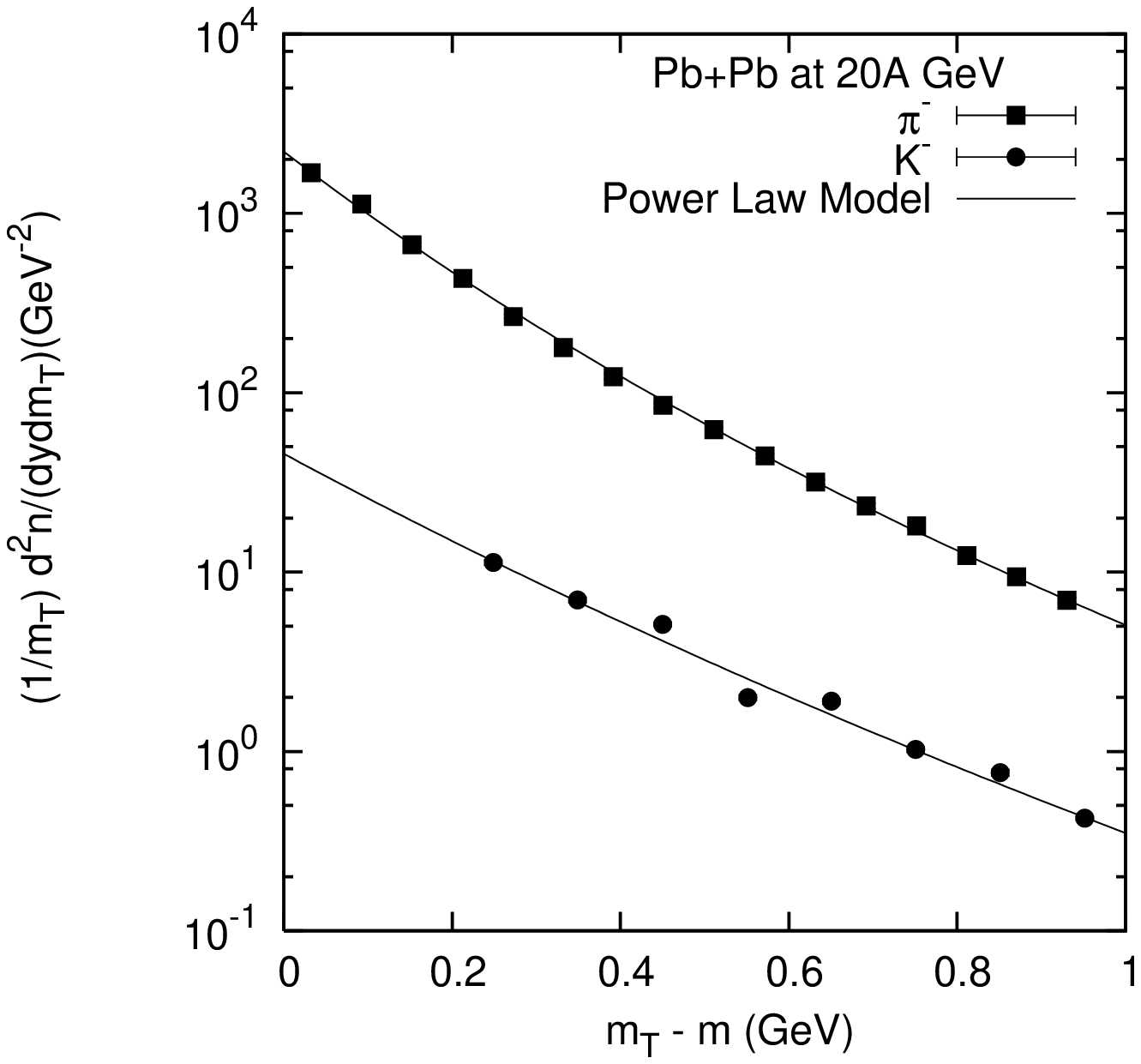}
 \end{minipage}}%
\caption{Transverse mass spectra of $\pi^+$, $K^+$, $p$ (left) and
$\pi^-$, $K^-$ (right) produced in central Pb+Pb
Collision at 20A GeV. The lines are fits of equation Power
Law Model. The statistical errors are smaller than the symbol
size, for which no errors are shown in the figure. Data are taken from Ref.\cite{Alt1} and \cite{Alt2}.}
\end{figure}

\begin{figure}
\subfigure[]{
\begin{minipage}{.5\textwidth}
\centering
 \includegraphics[width=2.5in]{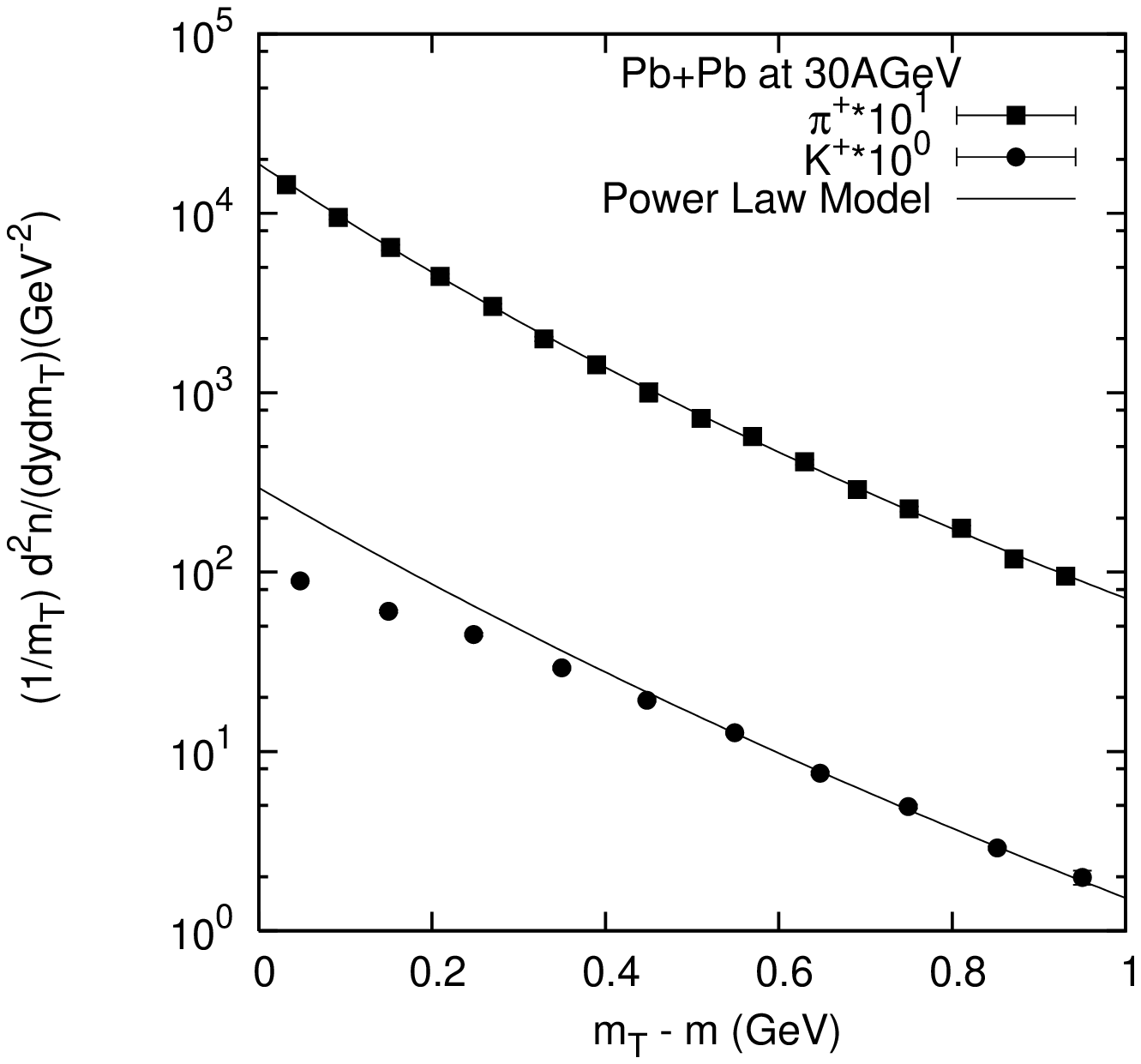}
\end{minipage}}%
\subfigure[]{
\begin{minipage}{0.5\textwidth}
  \centering
\includegraphics[width=2.5in]{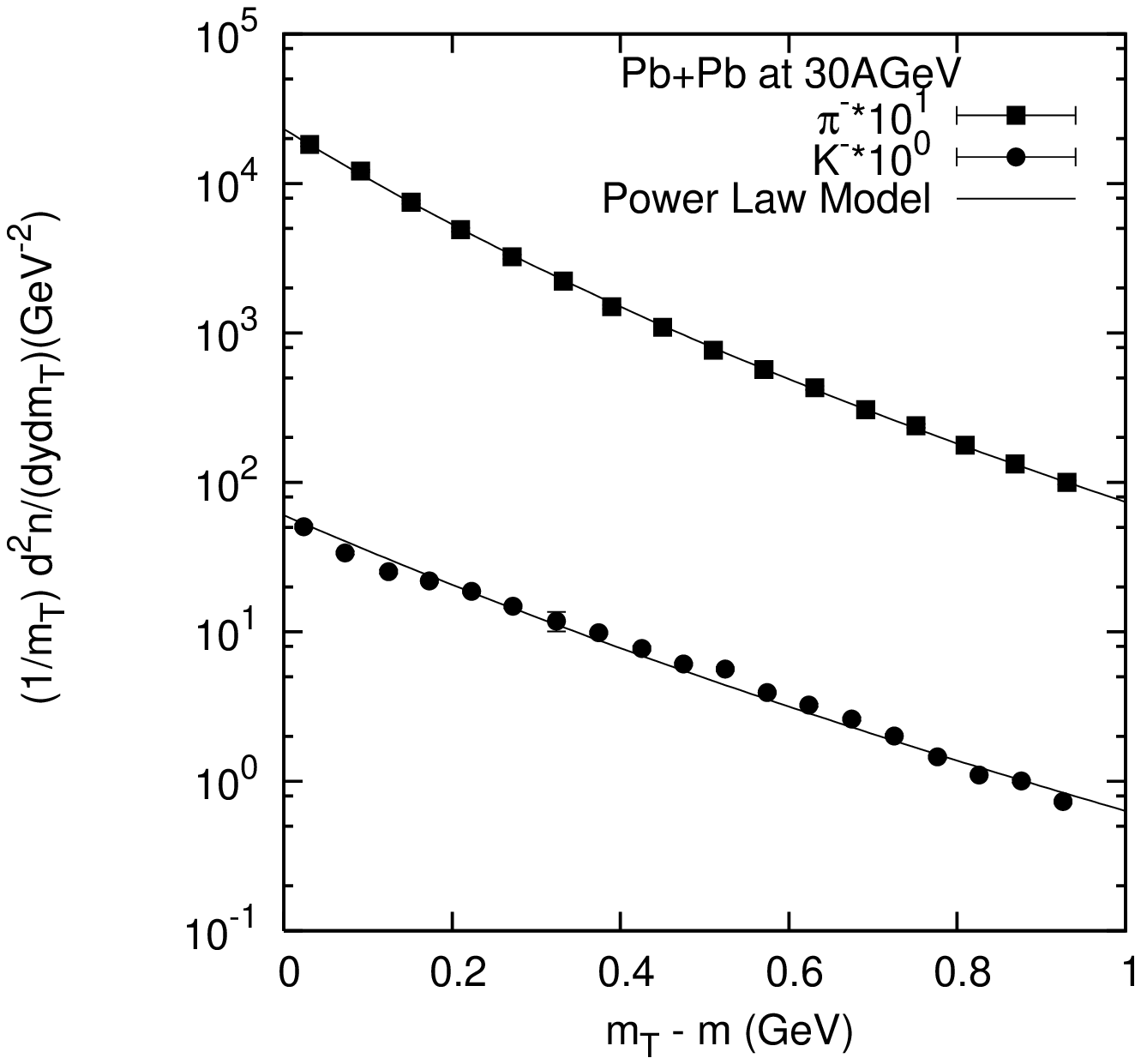}
\end{minipage}}%
\vspace{.01in} \subfigure[]{
\begin{minipage}{1\textwidth}
\centering
 \includegraphics[width=2.5in]{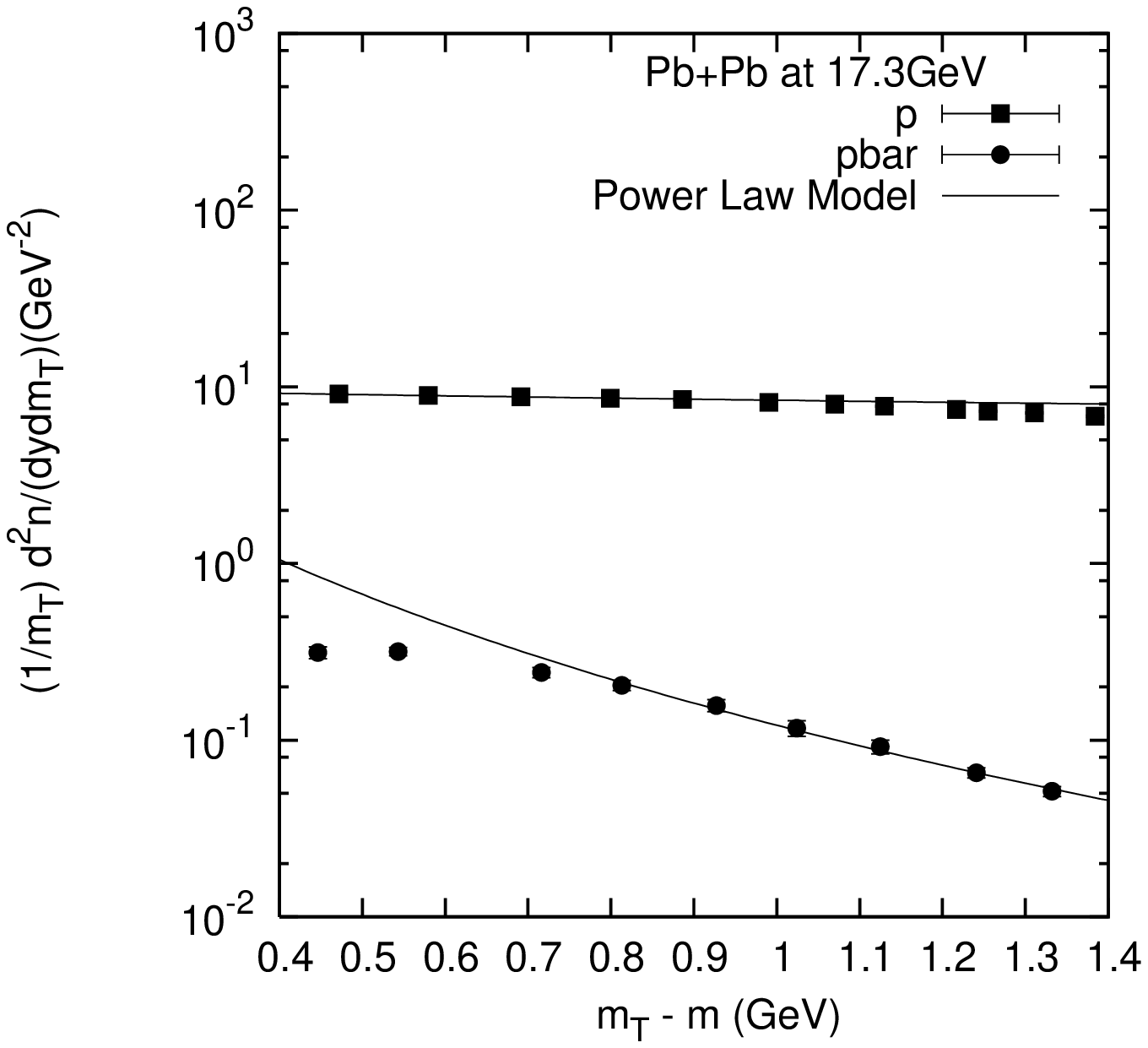}
\end{minipage}}%
\caption{Transverse mass spectra of $\pi^+$, $K^+$ (upper left) and  $\pi^-$, $K^-$
(upper right) and $p$, $\overline{p}$ (lower) produced in central Pb+Pb Collision at 30A  GeV
and 17.3 GeV. The lines are fits of equation of Power Law Model. The statistical errors are
smaller than the symbol size, for which no errors are shown in the figure. Data are taken from Ref.\cite{Alt1} and \cite{Alt3}.}
\end{figure}

\begin{figure}
\subfigure[]{
\begin{minipage}{.5\textwidth}
\centering
\includegraphics[width=2.5in]{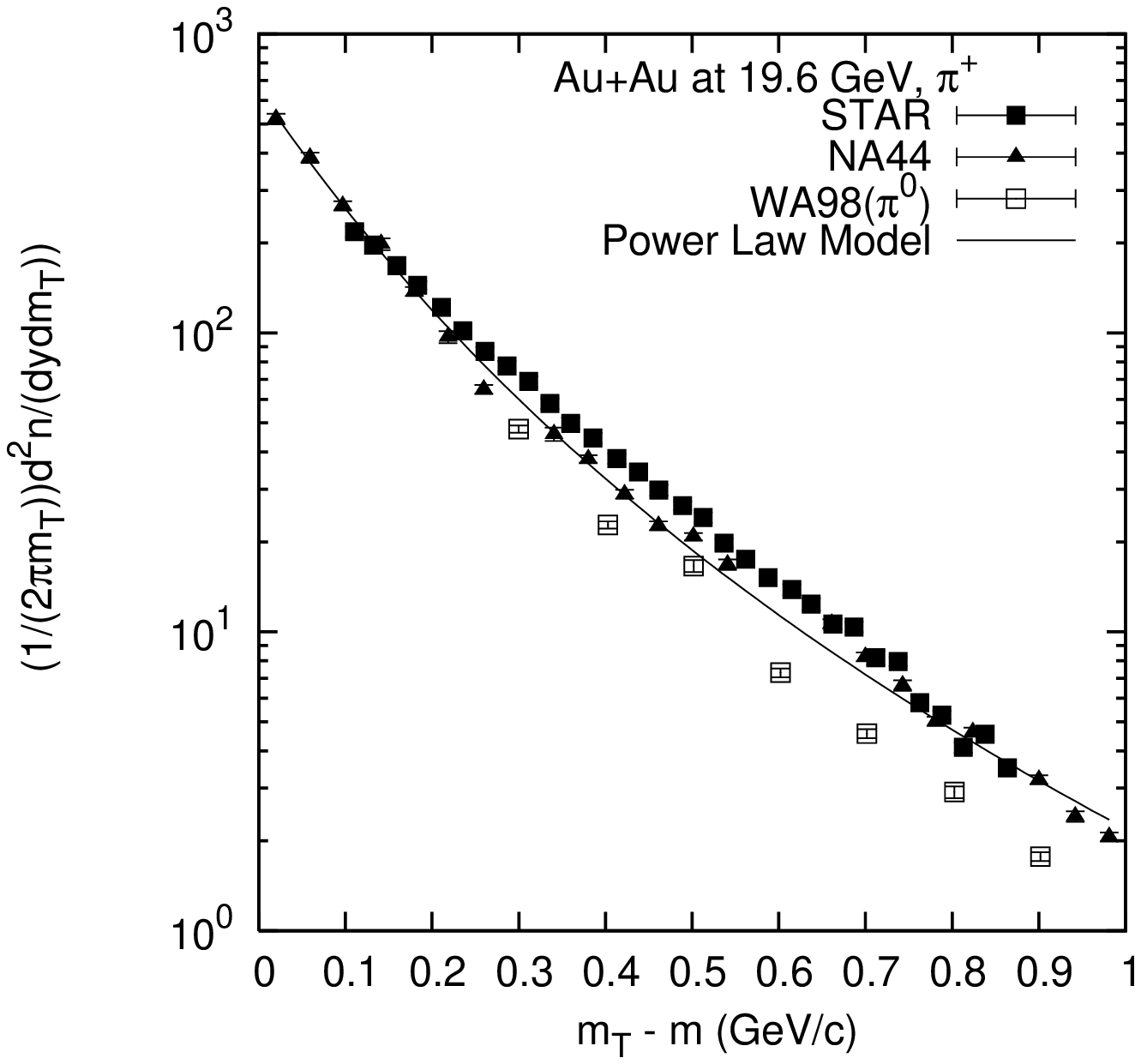}
\setcaptionwidth{2.6in}
\end{minipage}}%
\subfigure[]{
\begin{minipage}{0.5\textwidth}
\centering
 \includegraphics[width=2.5in]{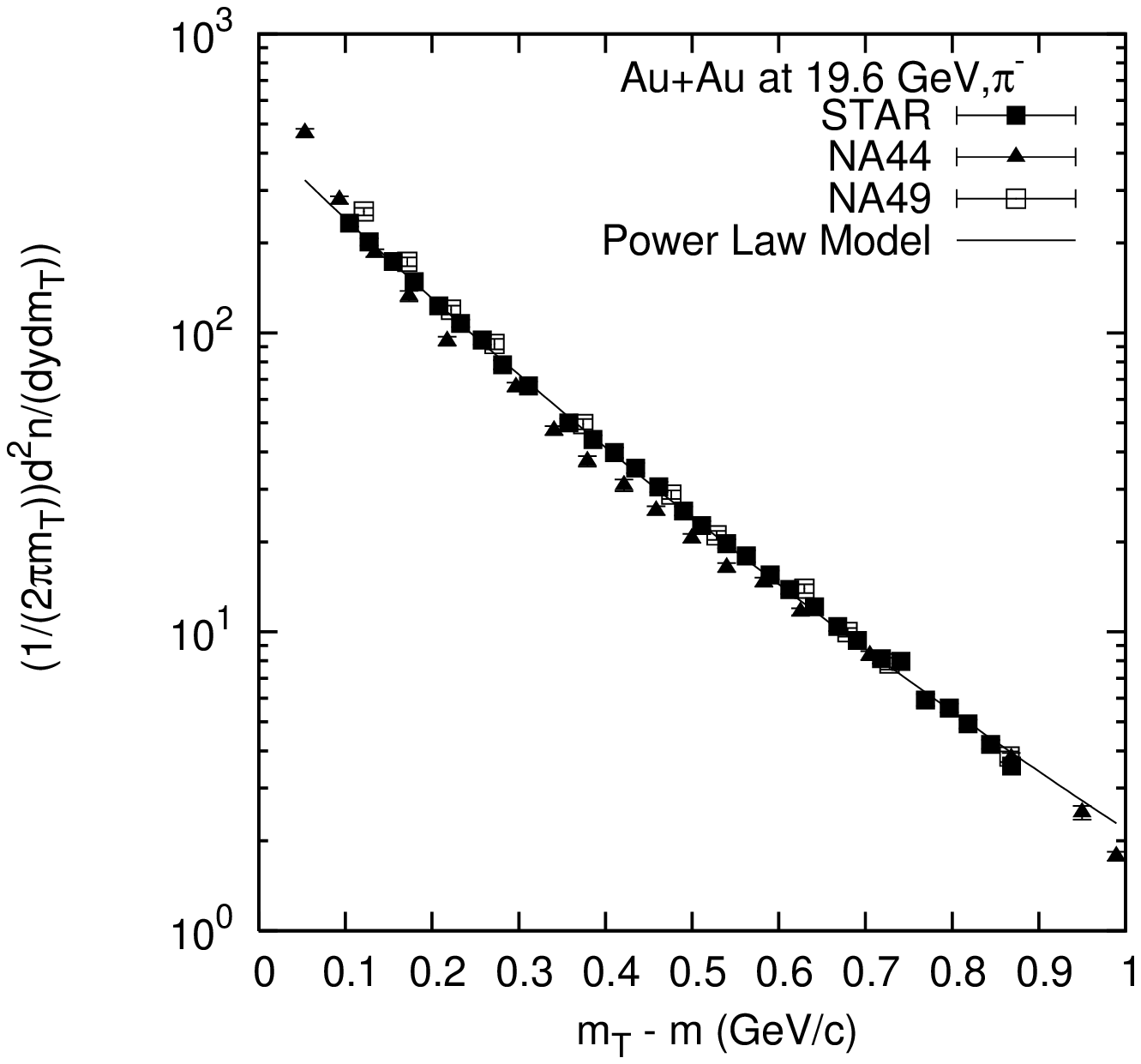}
 \end{minipage}}%
\vspace{0.01in} \subfigure[]{
\begin{minipage}{0.5\textwidth}
\centering
\includegraphics[width=2.5in]{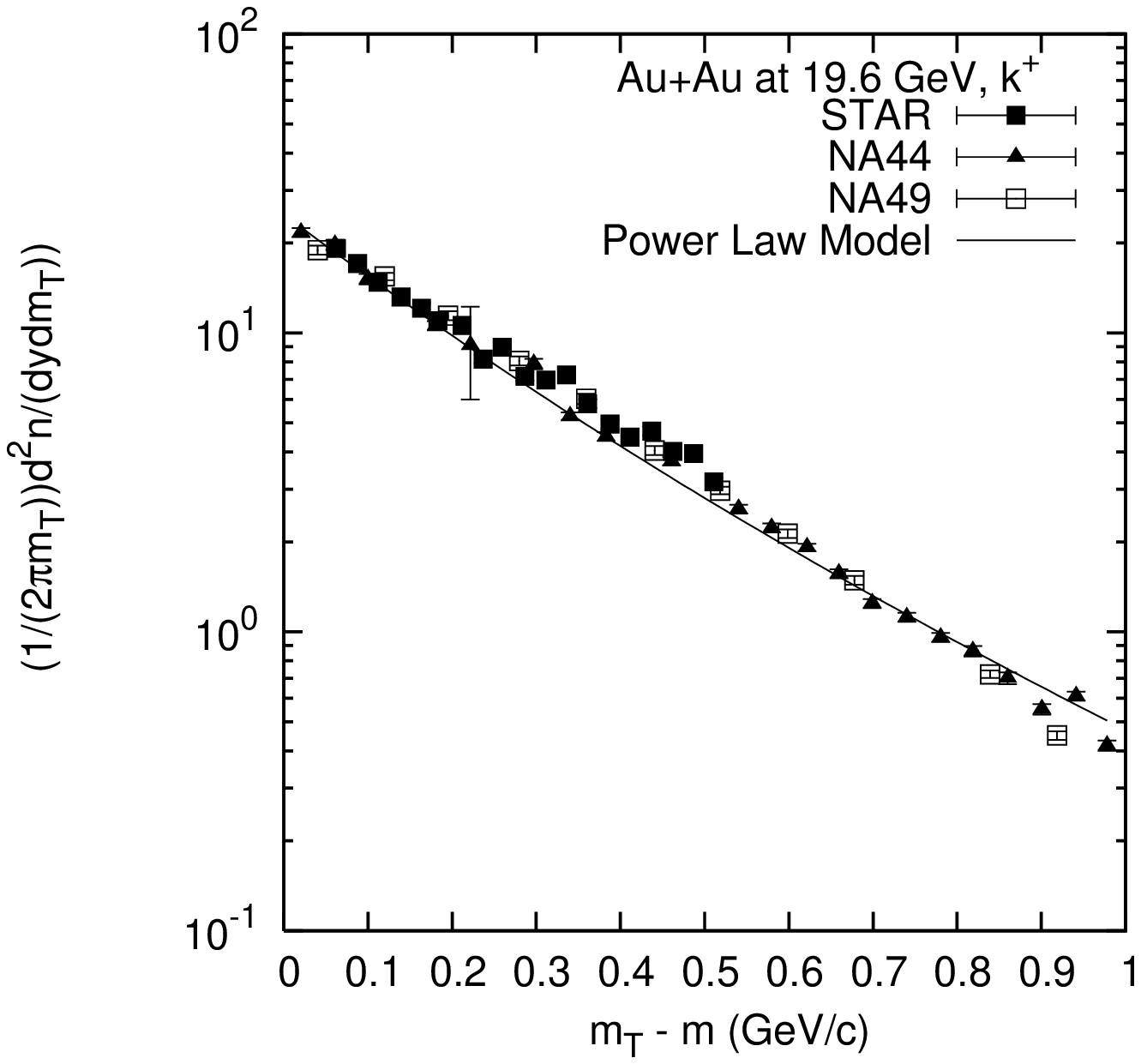}
\end{minipage}}%
\subfigure[]{
\begin{minipage}{.5\textwidth}
\centering
 \includegraphics[width=2.5in]{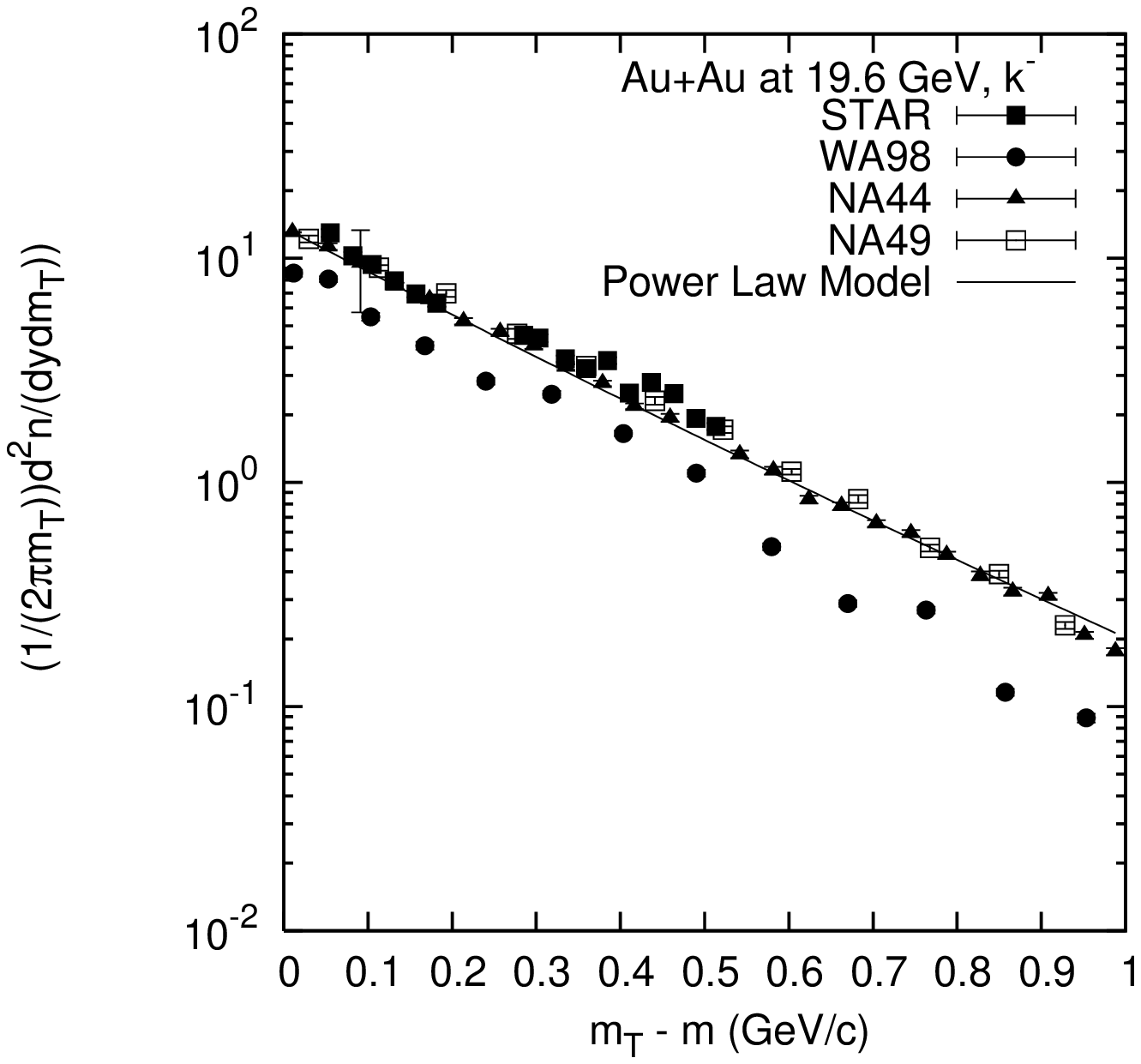}
 \end{minipage}}%
\caption{The transverse mass spectra of $\pi^+$ (upper left), $\pi^- $ (upper right)
and $K^+$ (lower left), $K^-$ (lower right)
from STAR experiment at 19.6 GeV in Au+Au collisions and the results
of SPS experiments NA44, NA49, WA98 at 17.3 GeV in Pb+Pb
collisions. The line is fit of Power Law Model with all the STAR and
SPS experiment. Data are taken from Ref.\cite{Cebra1} and all errors are only of statistical nature.}
\end{figure}

\begin{figure}
\subfigure[]{
\begin{minipage}{.5\textwidth}
\centering
\includegraphics[width=2.5in]{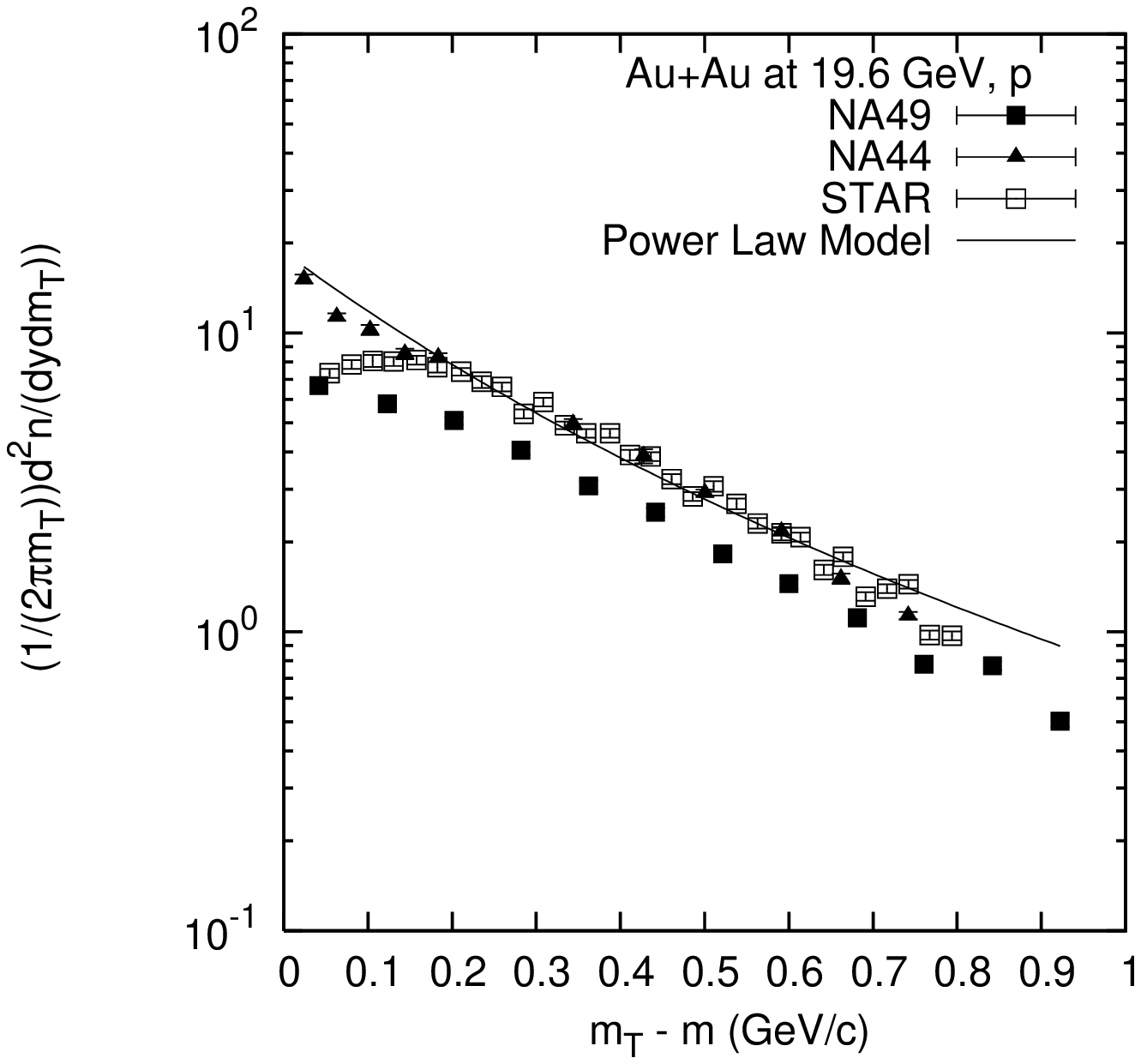}
\end{minipage}}%
\subfigure[]{
\begin{minipage}{.5\textwidth}
\centering
 \includegraphics[width=2.5in]{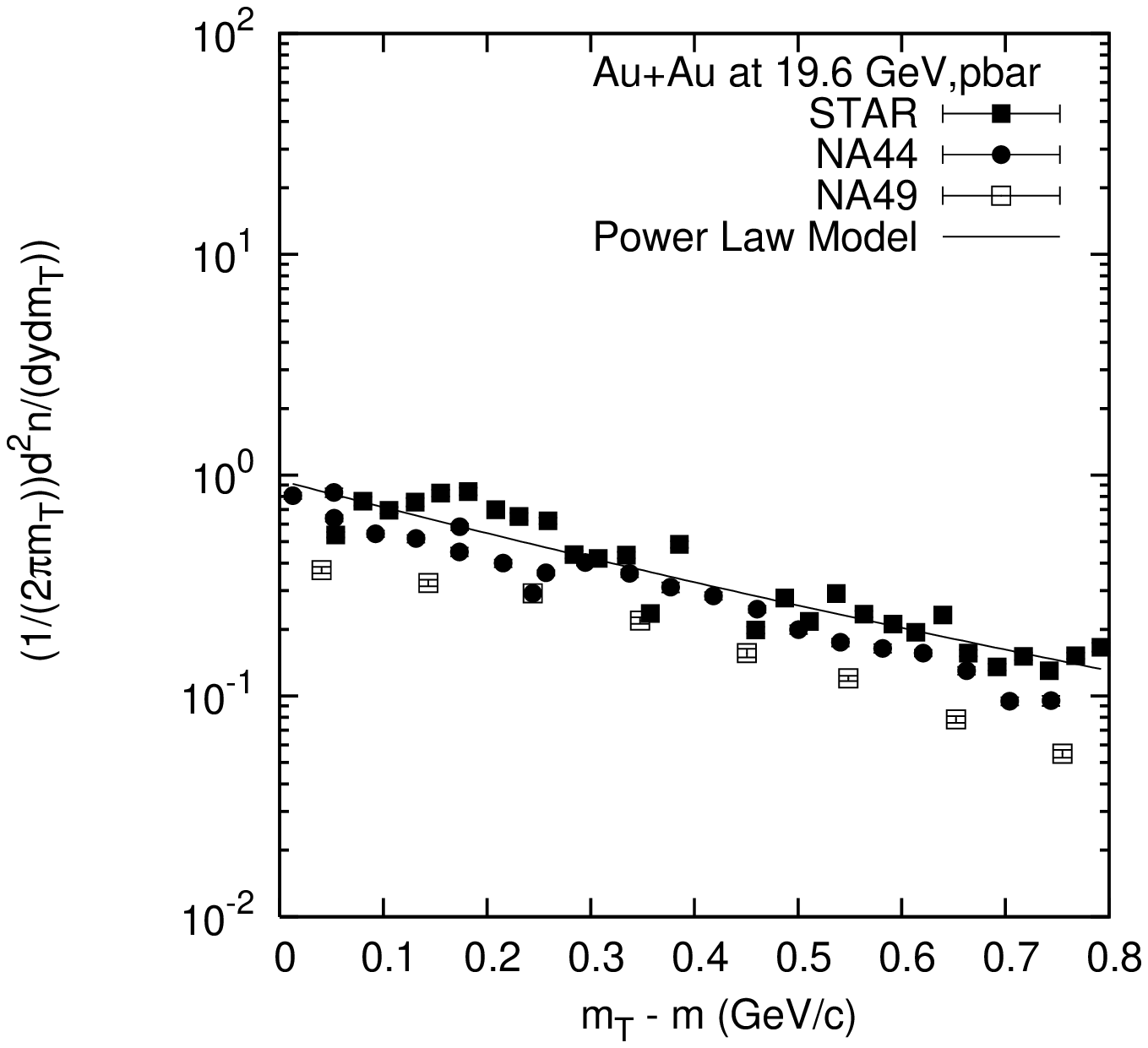}
 \end{minipage}}%
\caption{The transverse mass spectra of $p$ (left) and $\overline{p}$ (right) from STAR experiment
 at 19.6 GeV in Au+Au collisions and the results of SPS experiments NA44, NA49, WA98 at 17.3 GeV in
 Pb+Pb collisions. The line is fit of Power Law Model with all the STAR and SPS experiment.
 Data are taken from Ref.\cite{Cebra1}. The statistical errors are
smaller than the symbol size, for which no errors are shown in the figure. }
\end{figure}

\begin{figure}
\subfigure[]{
\begin{minipage}{.5\textwidth}
\centering
 \includegraphics[width=2.5in]{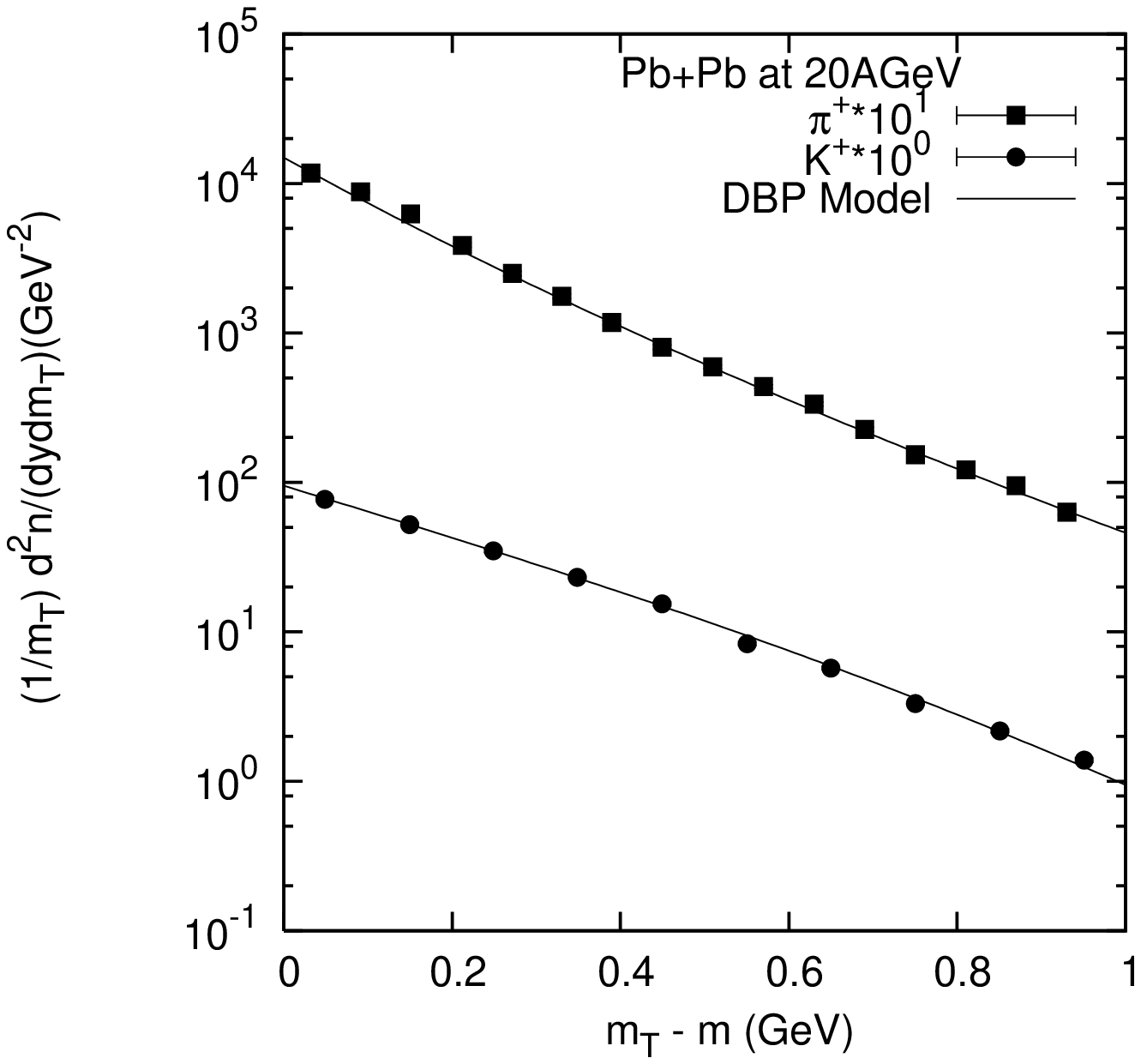}
\end{minipage}}%
\subfigure[]{
\begin{minipage}{0.5\textwidth}
  \centering
\includegraphics[width=2.5in]{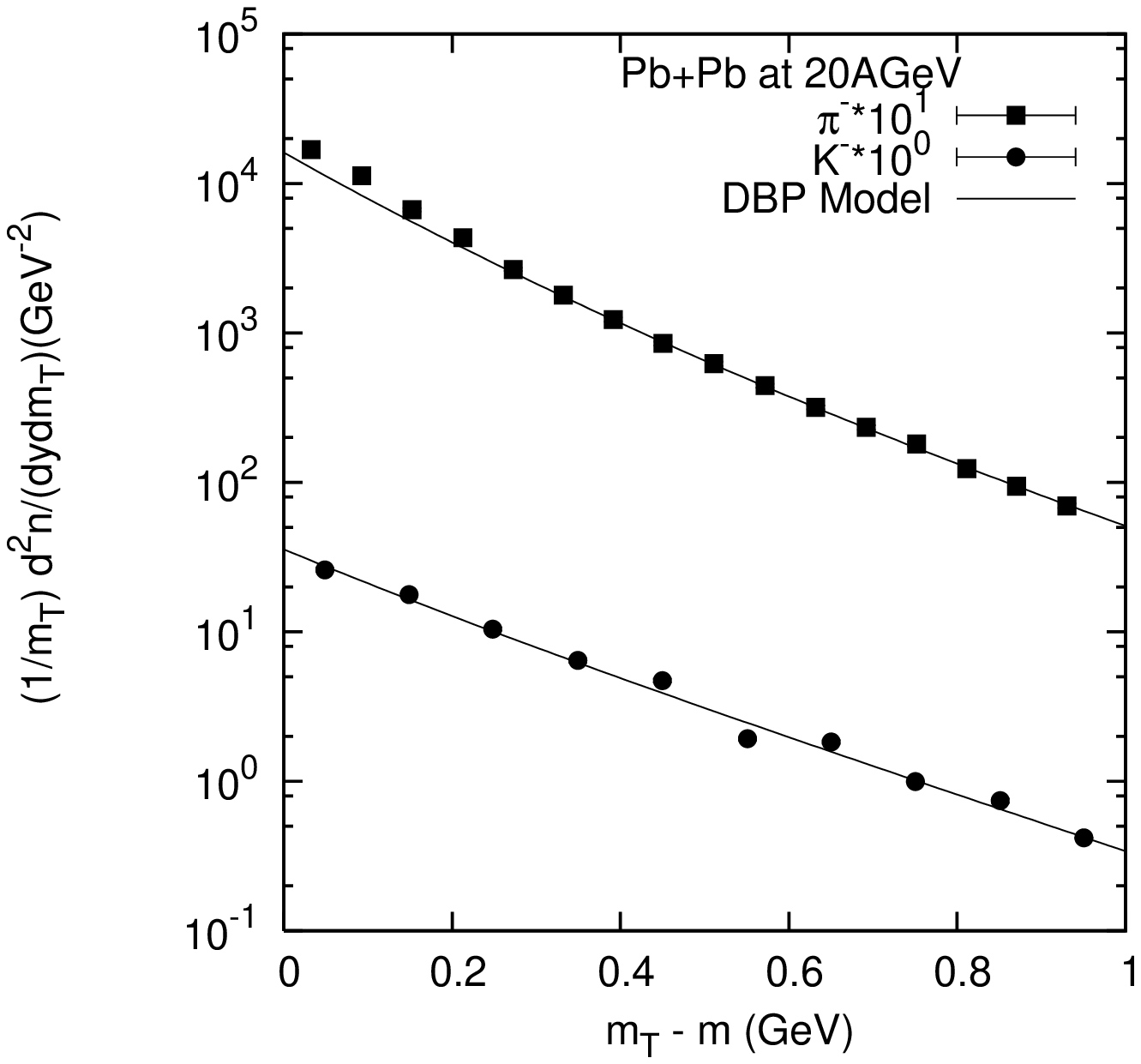}
\end{minipage}}%
\vspace{.01in} \subfigure[]{
\begin{minipage}{1\textwidth}
\centering
 \includegraphics[width=2.5in]{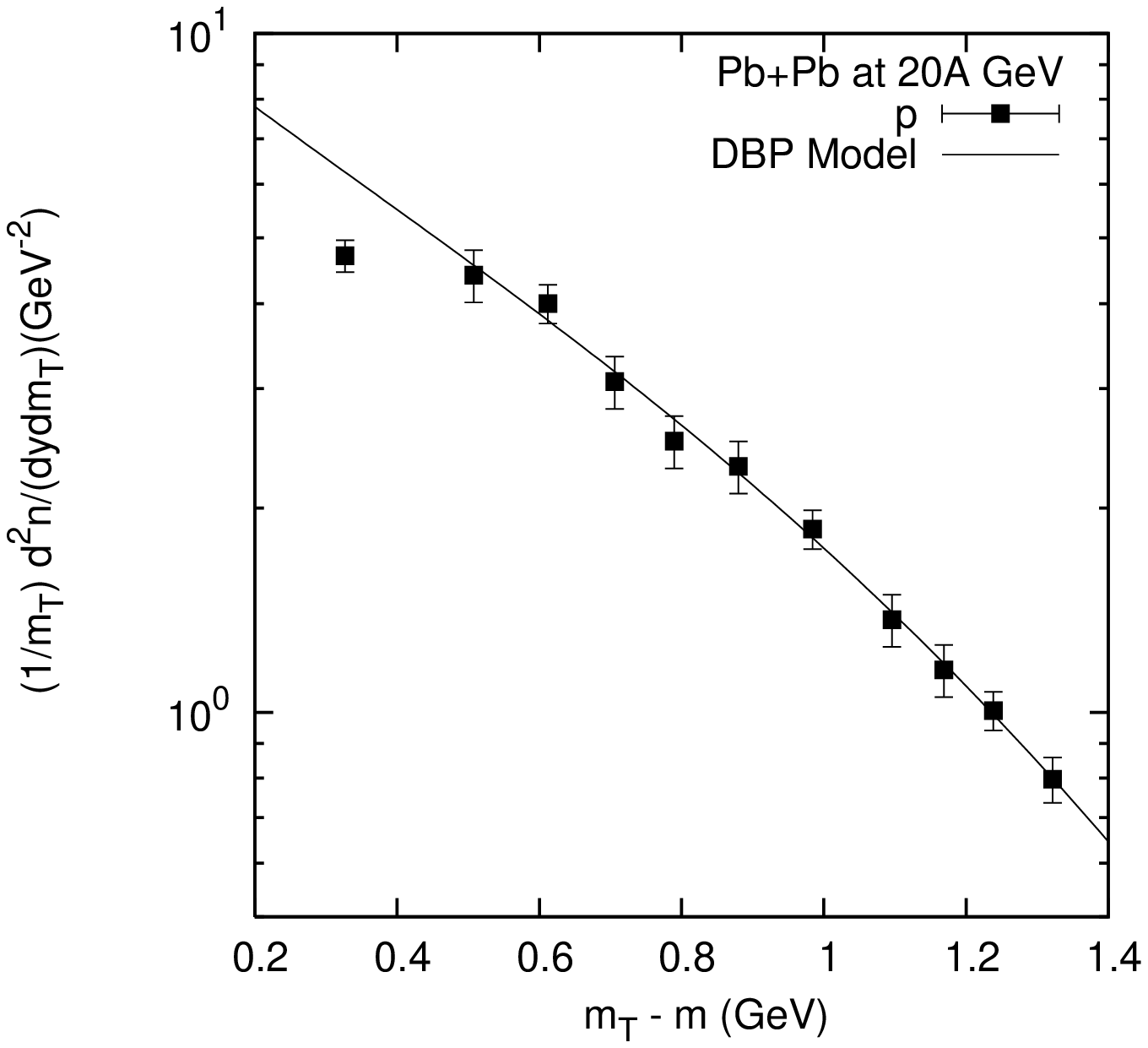}
\end{minipage}}%
\caption{Transverse mass spectra of $\pi^+$, $K^+$ (upper left) and $\pi^-$, $K^-$ (upper right)
and $p$ (lower) produced in central Pb+Pb Collision at 20A GeV.
 The lines are fits of equation of DBP Model. The statistical errors are smaller than the symbol
 size, for which no errors are shown in the figure. Data are taken from Ref.\cite{Alt1} and \cite{Alt2}. }
\end{figure}

\begin{figure}
\subfigure[]{
\begin{minipage}{.5\textwidth}
\centering
 \includegraphics[width=2.5in]{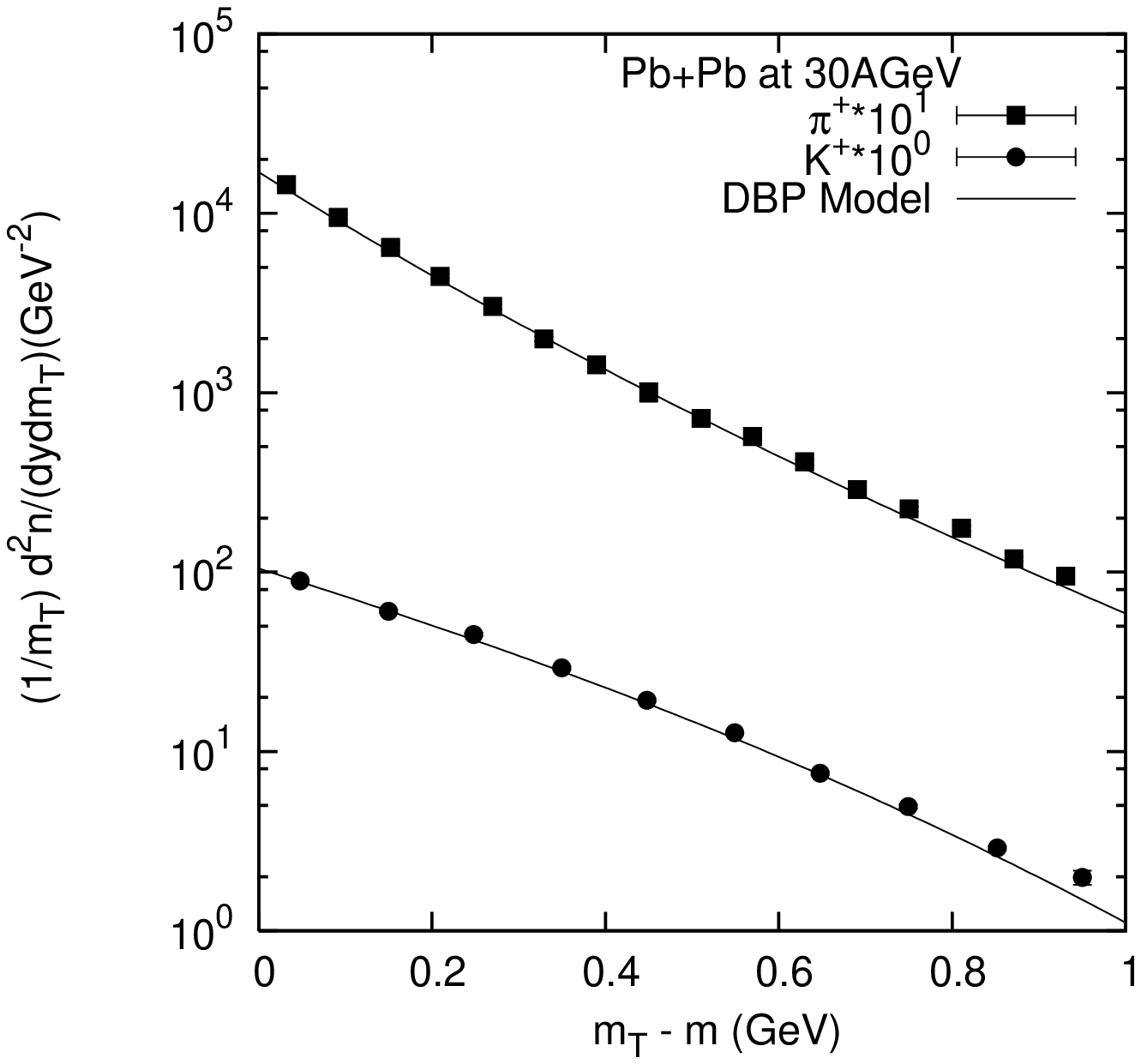}
\end{minipage}}%
\subfigure[]{
\begin{minipage}{0.5\textwidth}
  \centering
\includegraphics[width=2.5in]{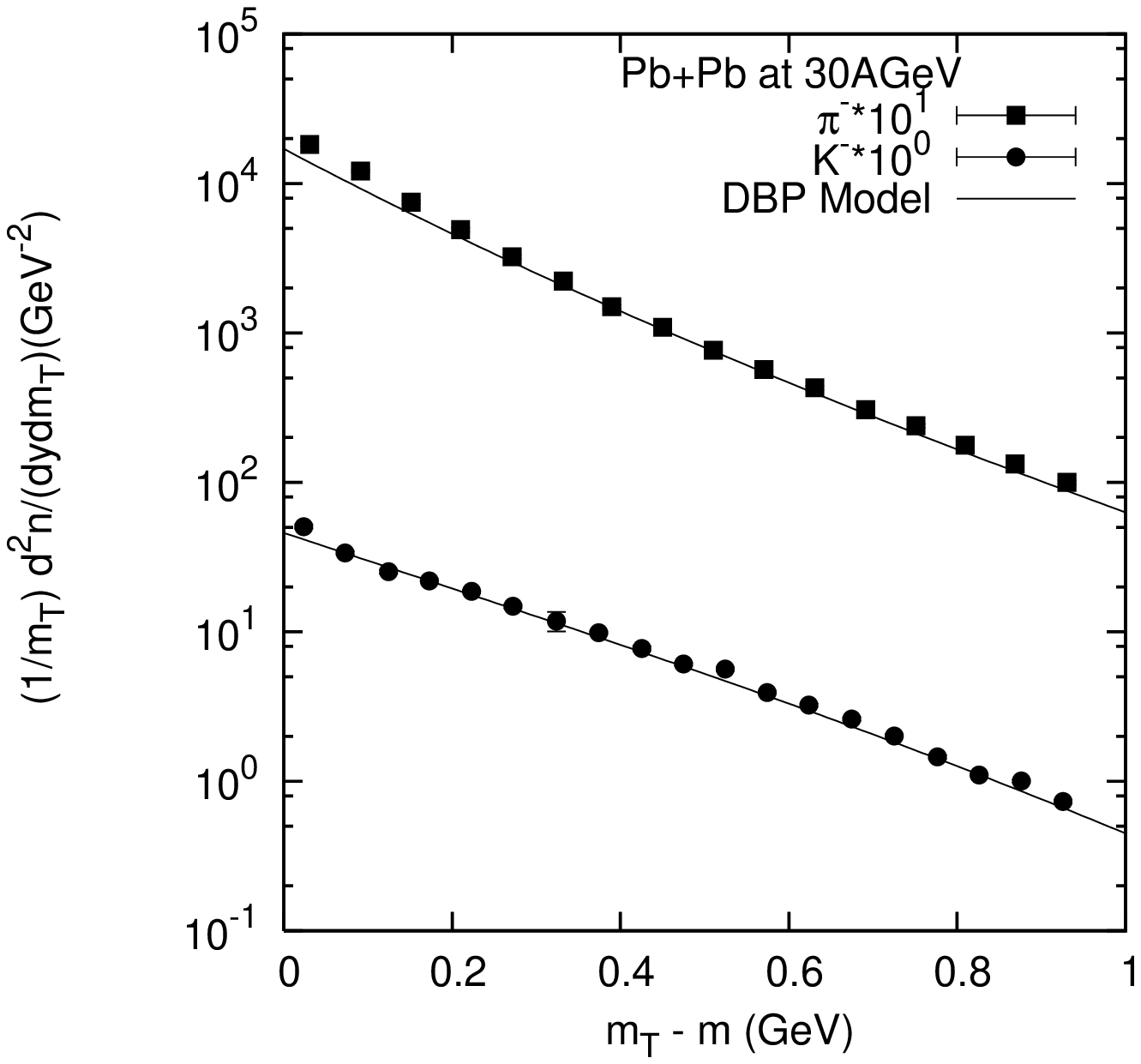}
\end{minipage}}%
\vspace{.01in} \subfigure[]{
\begin{minipage}{1\textwidth}
\centering
 \includegraphics[width=2.5in]{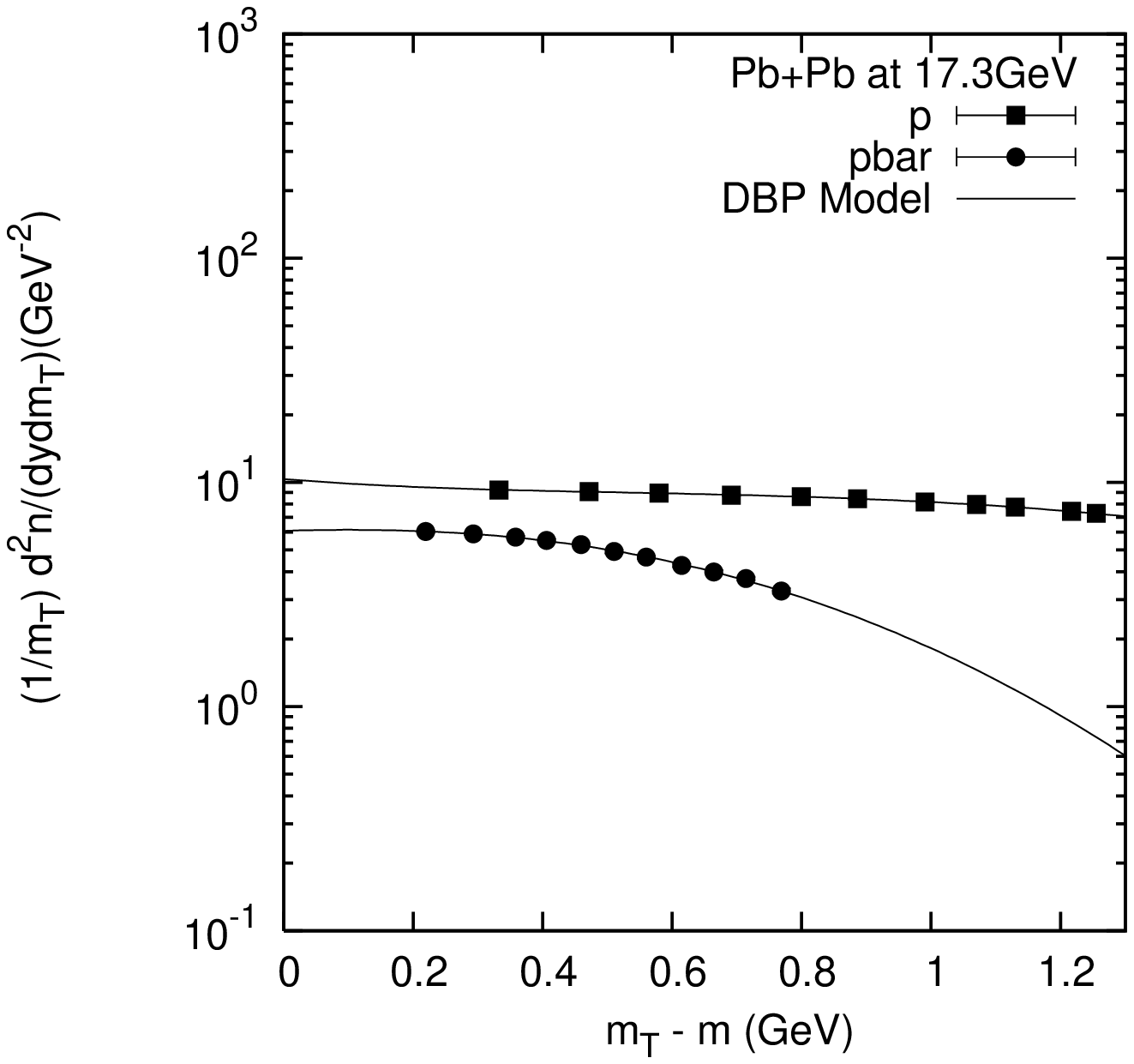}
\end{minipage}}%
\caption{Transverse mass spectra of $\pi^+$, $K^+$ (upper left) and $\pi^-$, $K^-$ (upper right)and $p$,
$\overline{p}$ (lower) produced in central Pb+Pb Collision at 30A  GeV and 17.3 GeV. The lines are
fits of equation of DBP Model. The statistical errors are smaller than the symbol size, for which no errors are
shown in the figure. Data are taken from Ref.\cite{Alt1} and \cite{Alt3}. }
\end{figure}

\begin{figure}
\subfigure[]{
\begin{minipage}{.5\textwidth}
\centering
\includegraphics[width=2.5in]{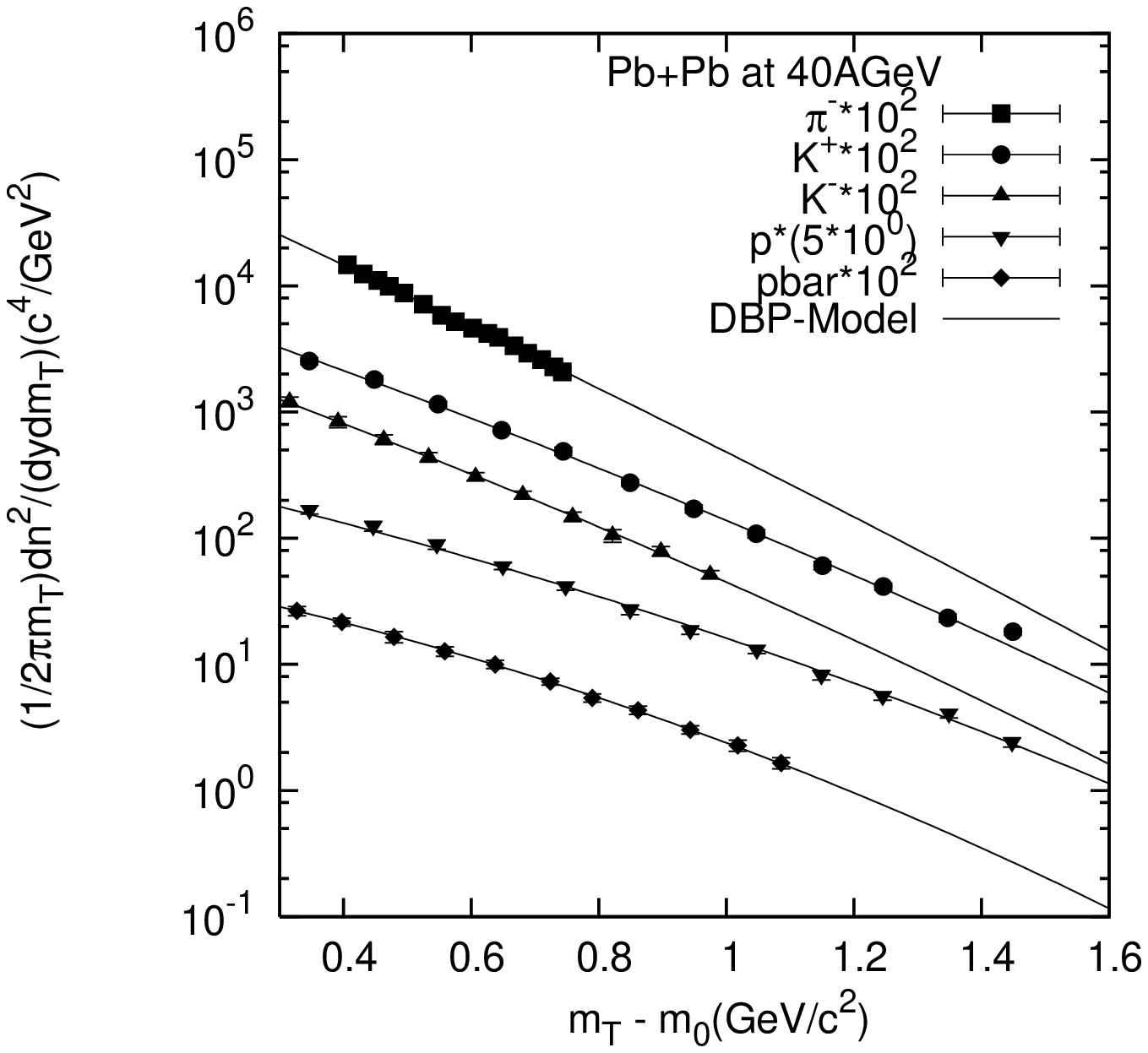}
\end{minipage}}%
\subfigure[]{
\begin{minipage}{.5\textwidth}
\centering
 \includegraphics[width=2.5in]{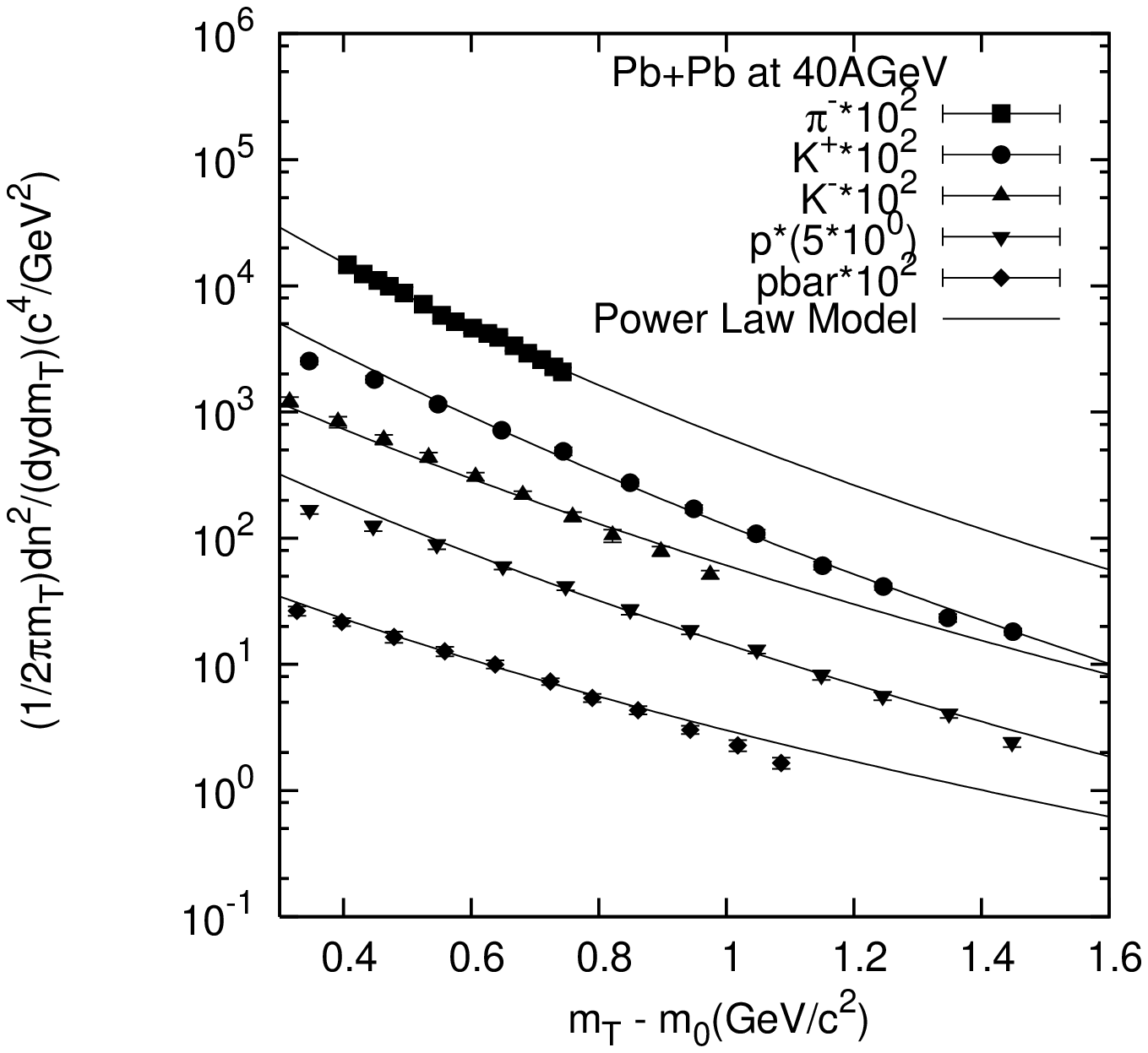}
 \end{minipage}}%
\caption{Transverse mass spectra of $K^+$, $K^-$, $p$, $\overline{p}$ and $\pi^-$ produced in
central Pb+Pb Collision at 40A GeV.
 The lines are fits of equation of DBP Model [Left Figure] and Power Law Model [Right Figure].
 The statistical errors are smaller than the symbol
 size, for which no errors are shown in the figure. Data are taken from \cite{Alt2}.}
\end{figure}

\begin{figure}
\subfigure[]{
\begin{minipage}{.5\textwidth}
\centering
\includegraphics[width=2.5in]{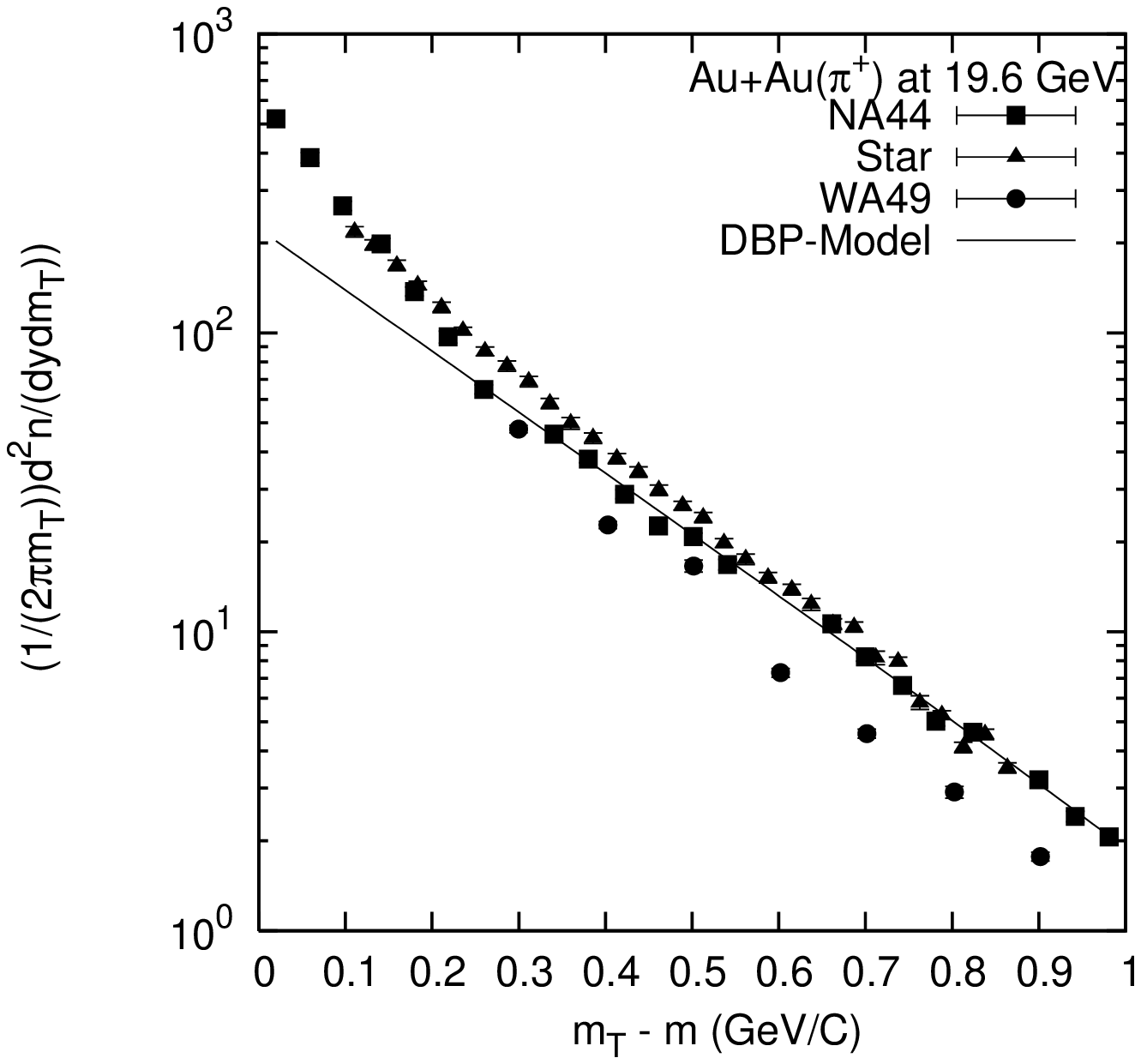}
\end{minipage}}%
\subfigure[]{
\begin{minipage}{.5\textwidth}
\centering
 \includegraphics[width=2.5in]{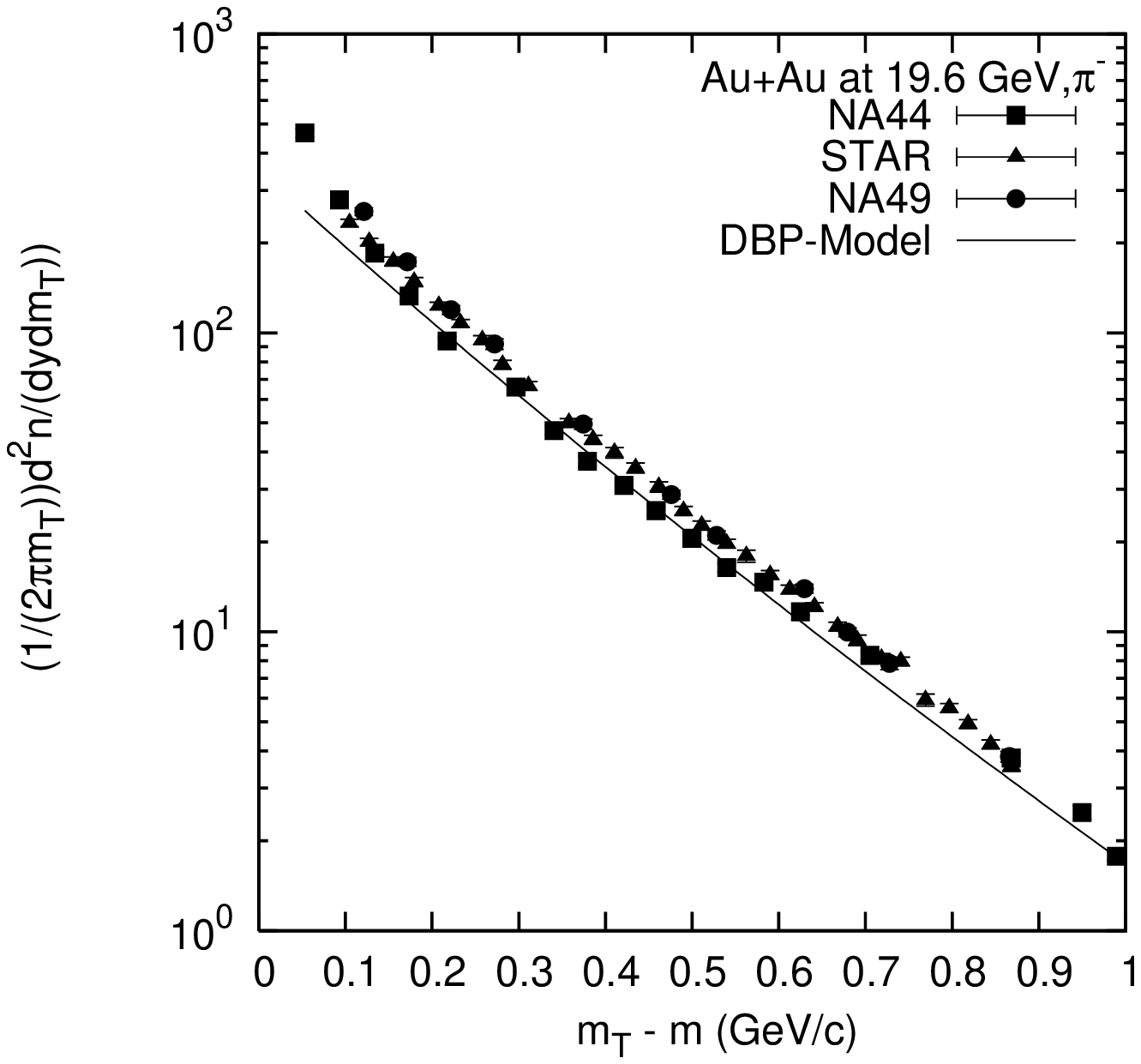}
 \end{minipage}}%
\caption{The transverse mass spectra of $\pi^+$ (left), $\pi^-$ (right) from STAR experiment at 19.6 GeV
in Au+Au collisions and the results of SPS experiments NA44, NA49, WA98 at 17.3 GeV in Pb+Pb collisions. The
line is fit of Power Law Model with all the STAR and SPS experiment. Data are taken from Ref.\cite{Cebra1}. The statistical
errors are smaller than the symbol size, for which no errors are shown in the figure. }
\end{figure}

\begin{figure}
\subfigure[]{
\begin{minipage}{.5\textwidth}
\centering
\includegraphics[width=2.5in]{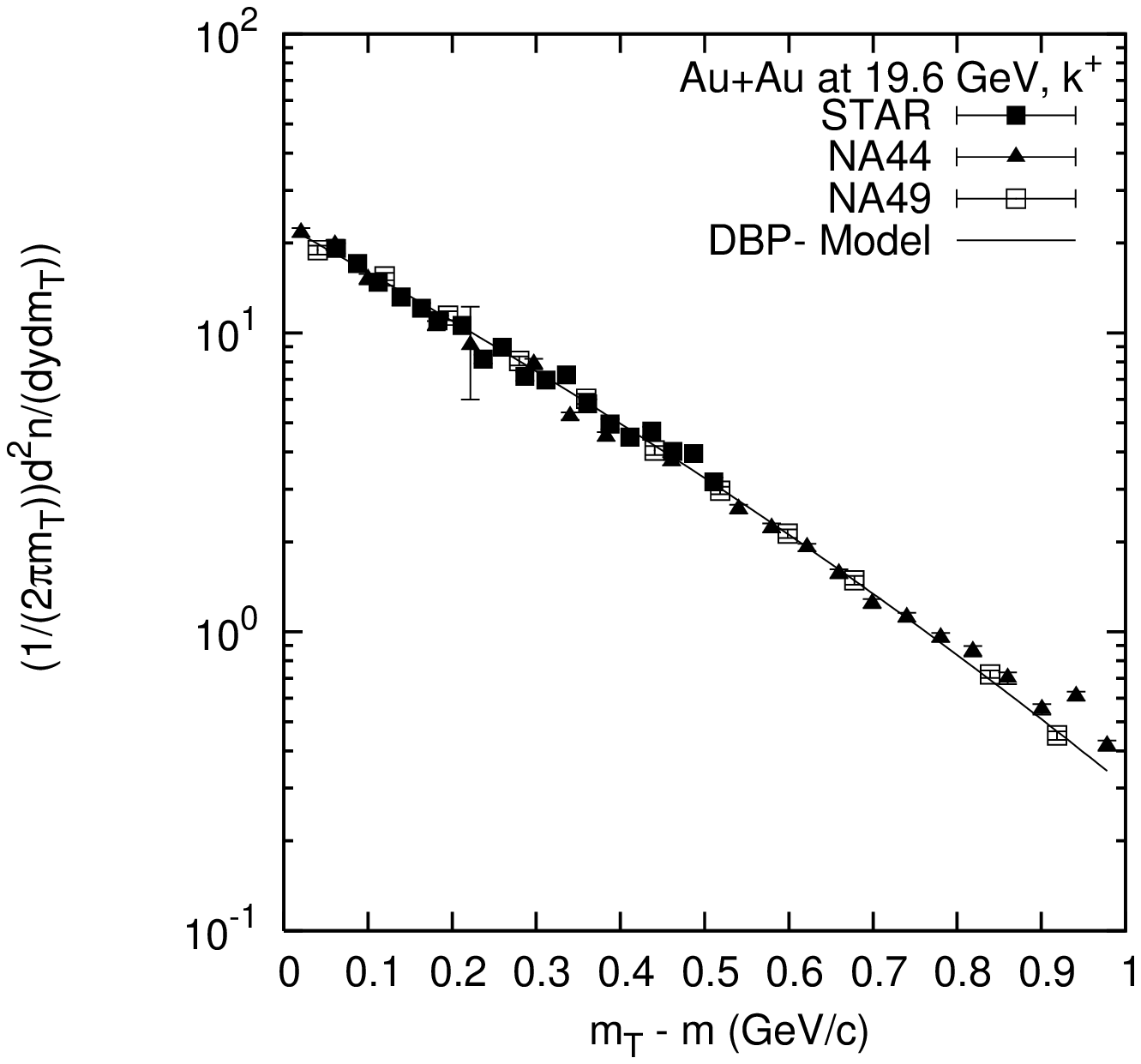}
\setcaptionwidth{2.6in}
\end{minipage}}%
\subfigure[]{
\begin{minipage}{0.5\textwidth}
\centering
 \includegraphics[width=2.5in]{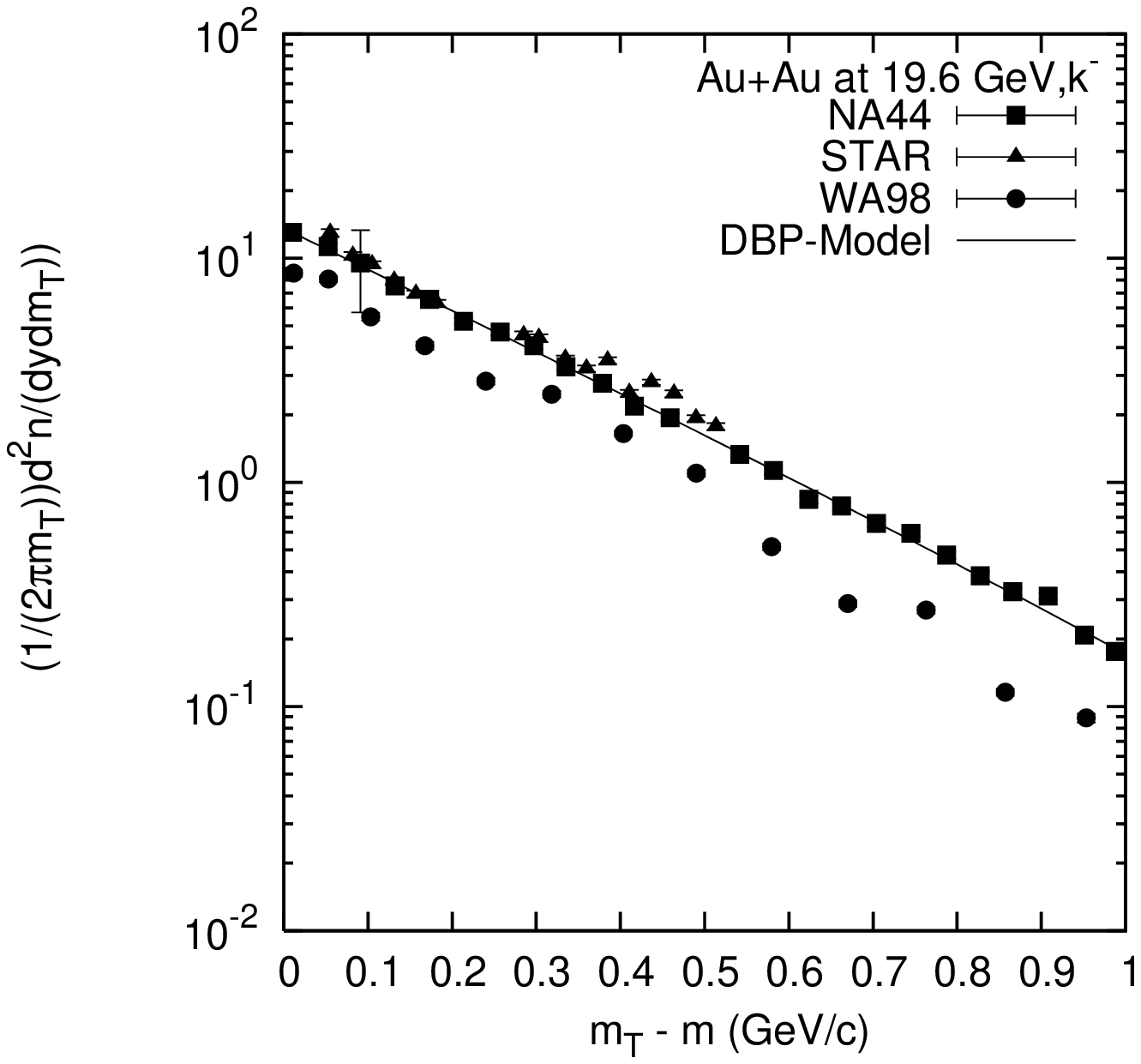}
 \end{minipage}}%
\vspace{0.01in} \subfigure[]{
\begin{minipage}{0.5\textwidth}
\centering
\includegraphics[width=2.5in]{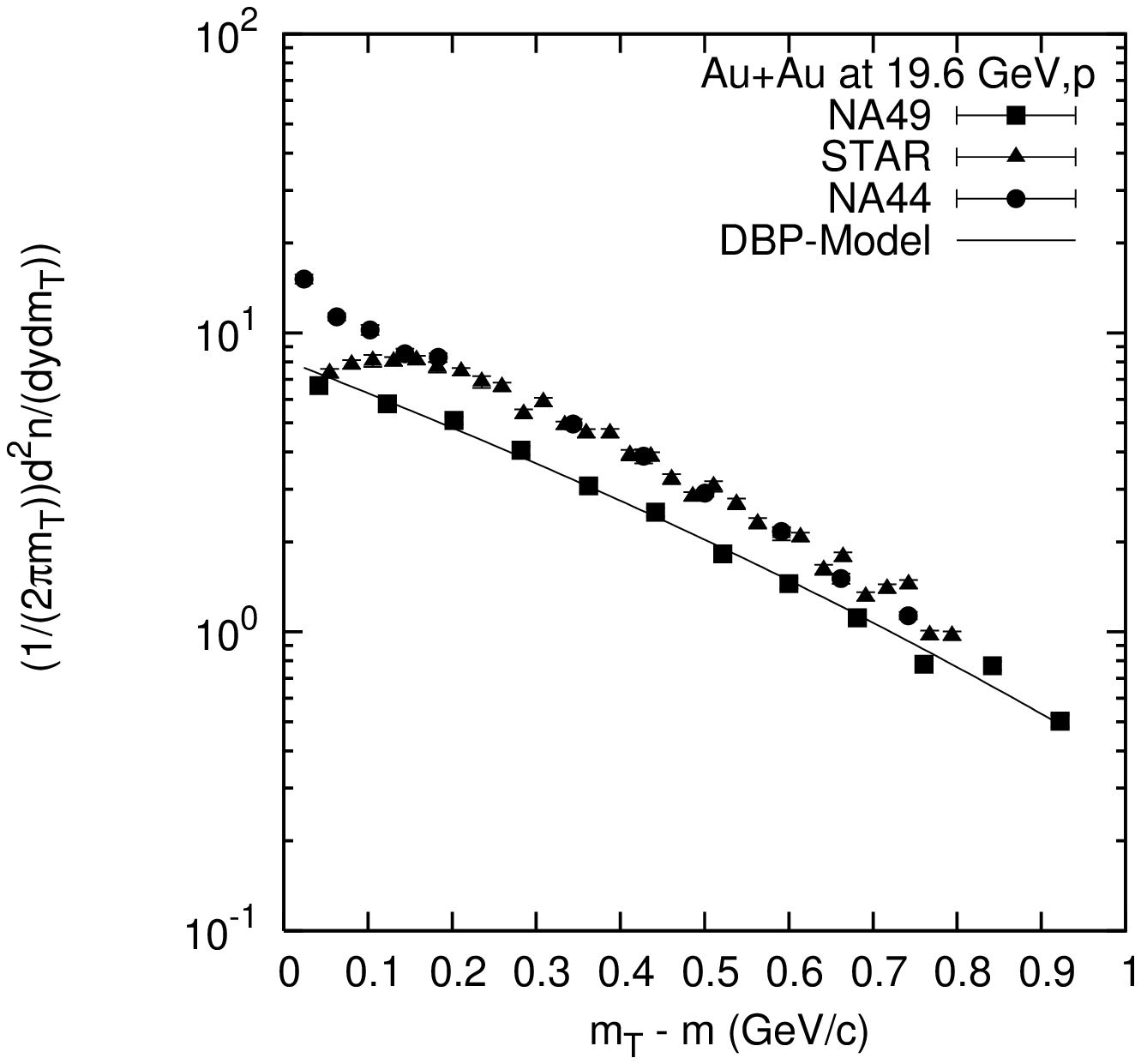}
\end{minipage}}%
\subfigure[]{
\begin{minipage}{.5\textwidth}
\centering
 \includegraphics[width=2.5in]{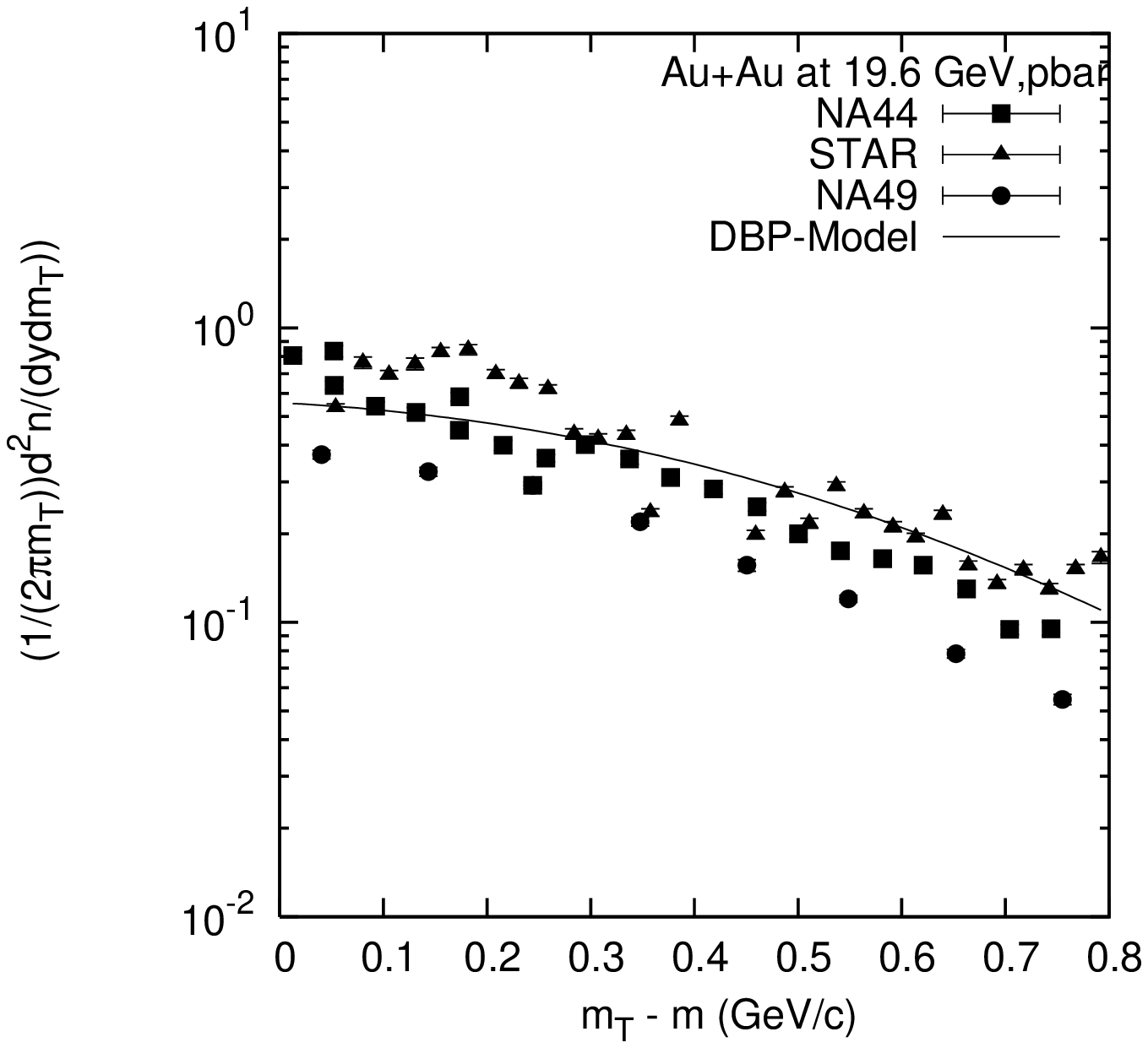}
 \end{minipage}}%
\caption{The transverse mass spectra of $K^+$ (upper left), $K^-$ (upper right) and $p$ (lower left),
$\overline{p}$ (lower right) from STAR experiment at 19.6 GeV in Au+Au collisions and the results of SPS
experiments NA44, NA49, WA98 at 17.3 GeV in Pb+Pb collisions. The line is fit of Power Law Model with all the STAR
and SPS experiment. Data are taken from Ref.\cite{Cebra1}. All errors are only of statistical nature.}
\end{figure}

\begin{figure}
\subfigure[]{
\begin{minipage}{.5\textwidth}
\centering
\includegraphics[width=2.5in]{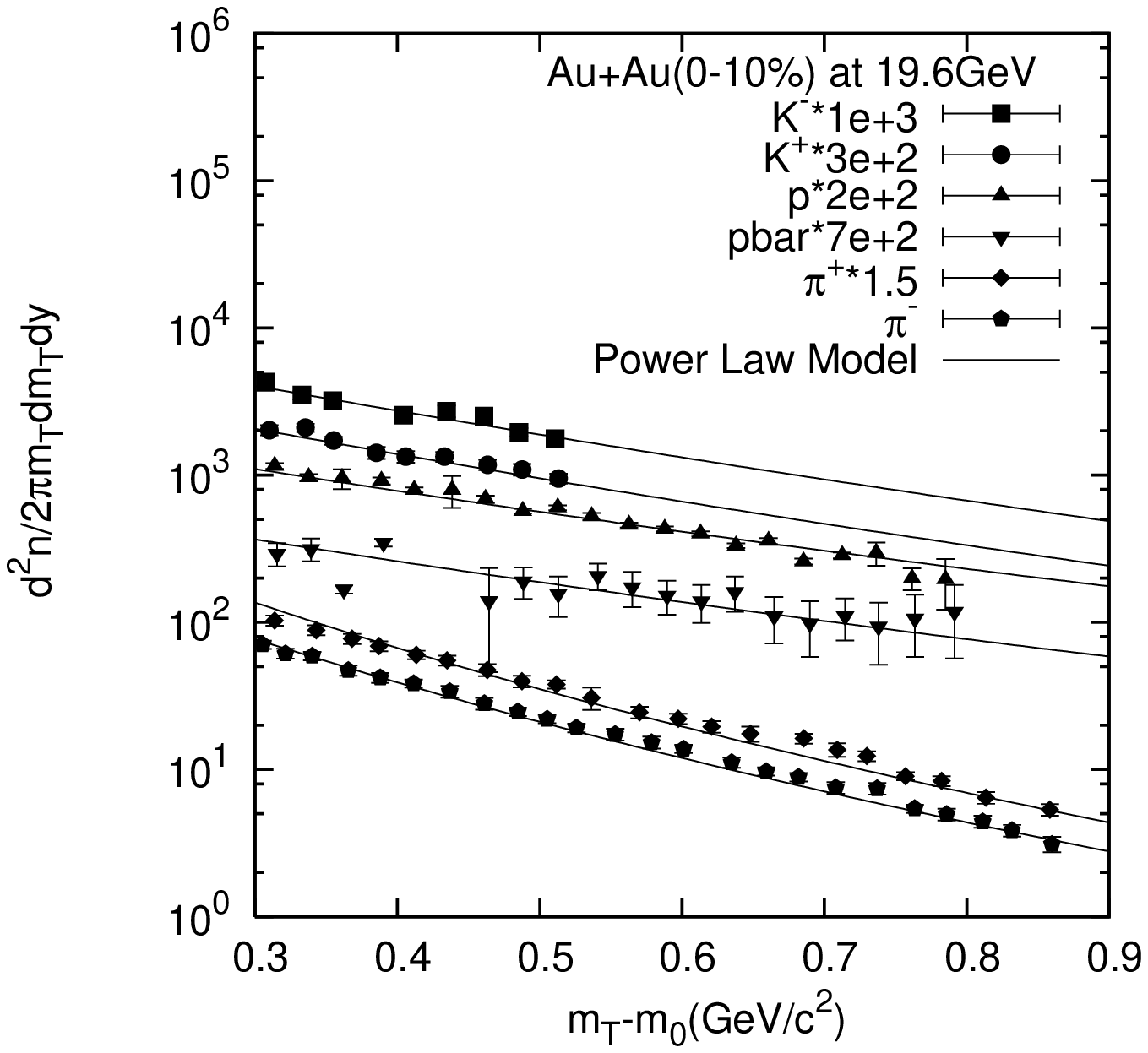}
\setcaptionwidth{2.6in}
\end{minipage}}%
\subfigure[]{
\begin{minipage}{0.5\textwidth}
\centering
 \includegraphics[width=2.5in]{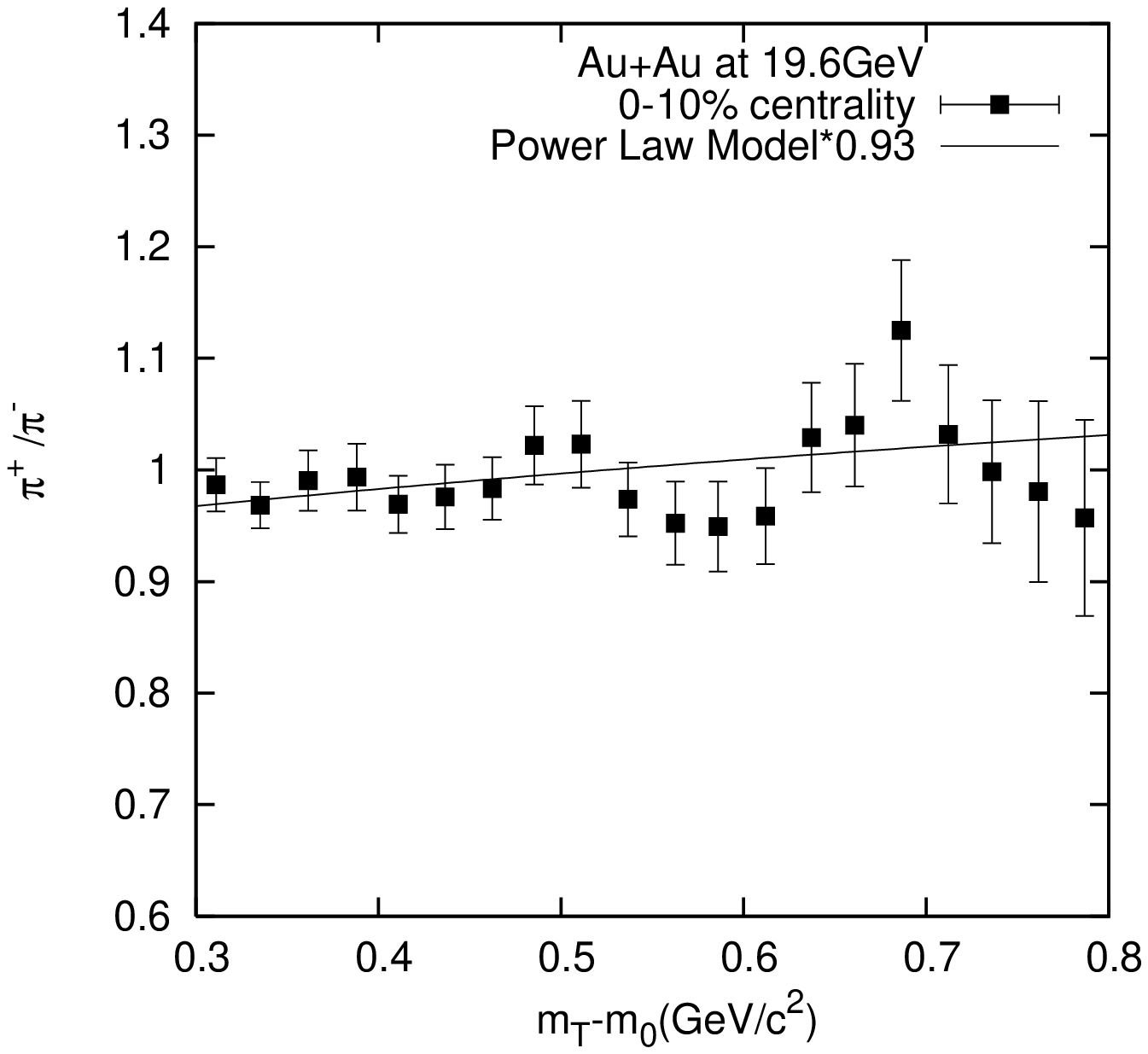}
 \end{minipage}}%
\vspace{0.01in} \subfigure[]{
\begin{minipage}{0.5\textwidth}
\centering
\includegraphics[width=2.5in]{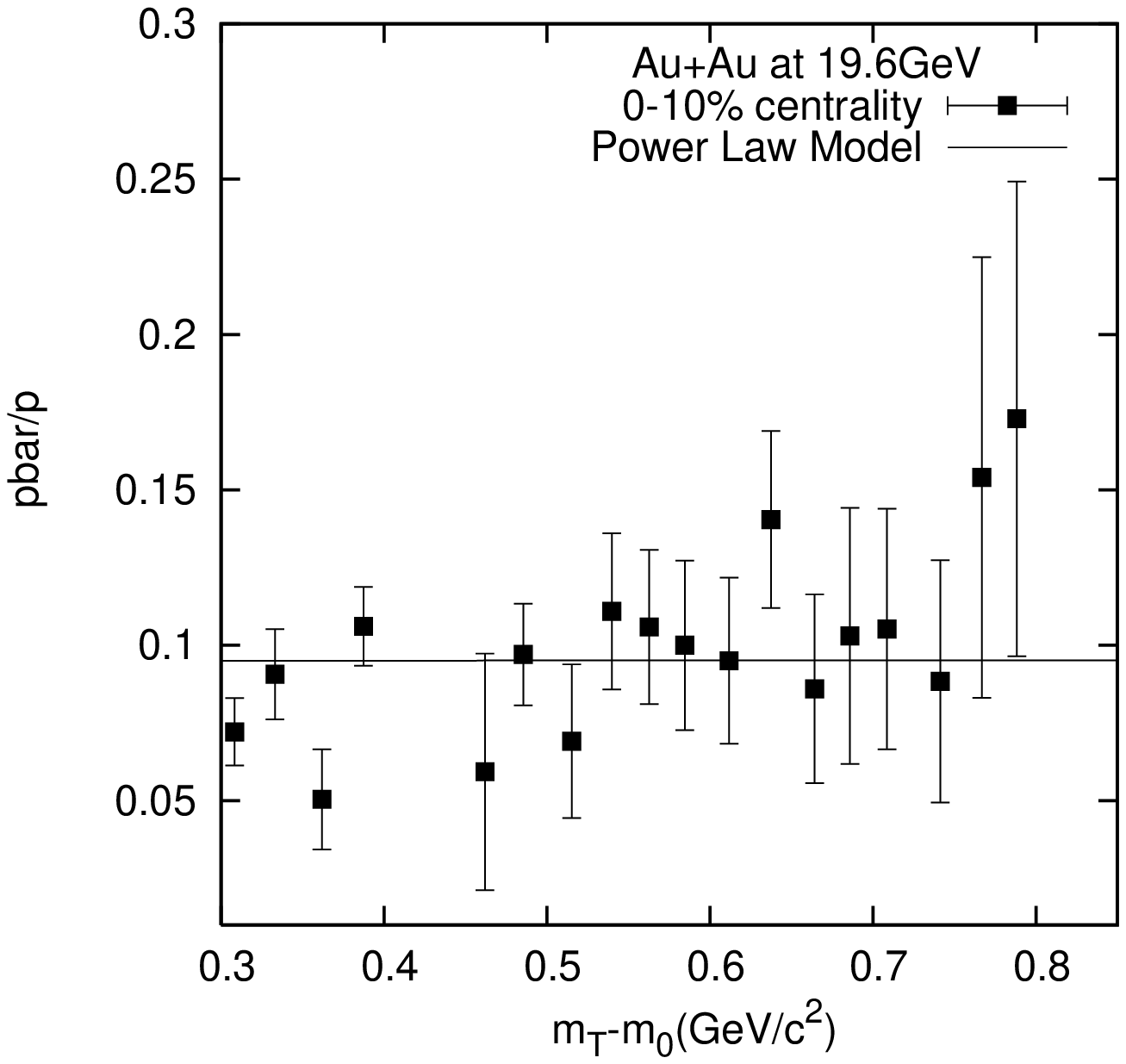}
\end{minipage}}%
\subfigure[]{
\begin{minipage}{.5\textwidth}
\centering
 \includegraphics[width=2.5in]{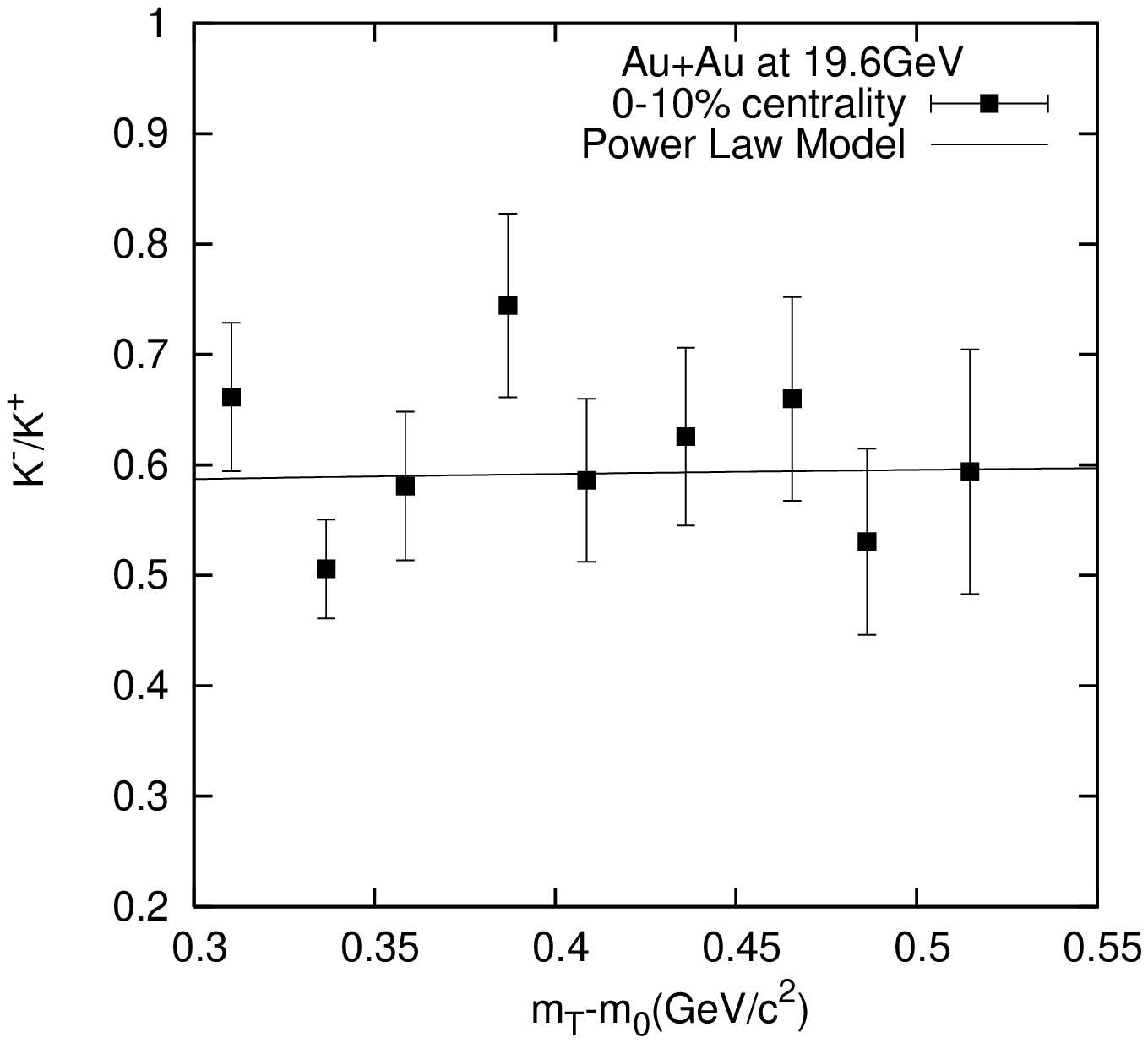}
 \end{minipage}}%
\caption{(a) Transverse mass spectra of identified hadrons measured at midrapidity $(|y|<0.1)$. The results at
$\sqrt{s_{NN}}$=19.6 GeV for the production of $\pi^+$, $\pi^-$, p, $\overline{p}$, $K^+$ and $K^-$
for 0-10\% centrality in Au+Au collisions. The solid curves provide the Power Law Model based results. Data are taken from Ref.\cite{Picha1}
(b),(c),(d) $\pi^-/\pi^+$, $\overline{p}/p$ and $K^-/K^+$ ratios vs. $m_T-m_0$ for 0-10\% centrality in Au+Au collisions at
 $\sqrt{s_{NN}}$=19.6 GeV $(-0.1<y<0.1)$. The solid curves provide the Power Law Model based results. Data are taken from Ref.\cite{Picha1}.
All errors are only of statistical nature.}
\end{figure}

\begin{figure}
\subfigure[]{
\begin{minipage}{.5\textwidth}
\centering
\includegraphics[width=2.5in]{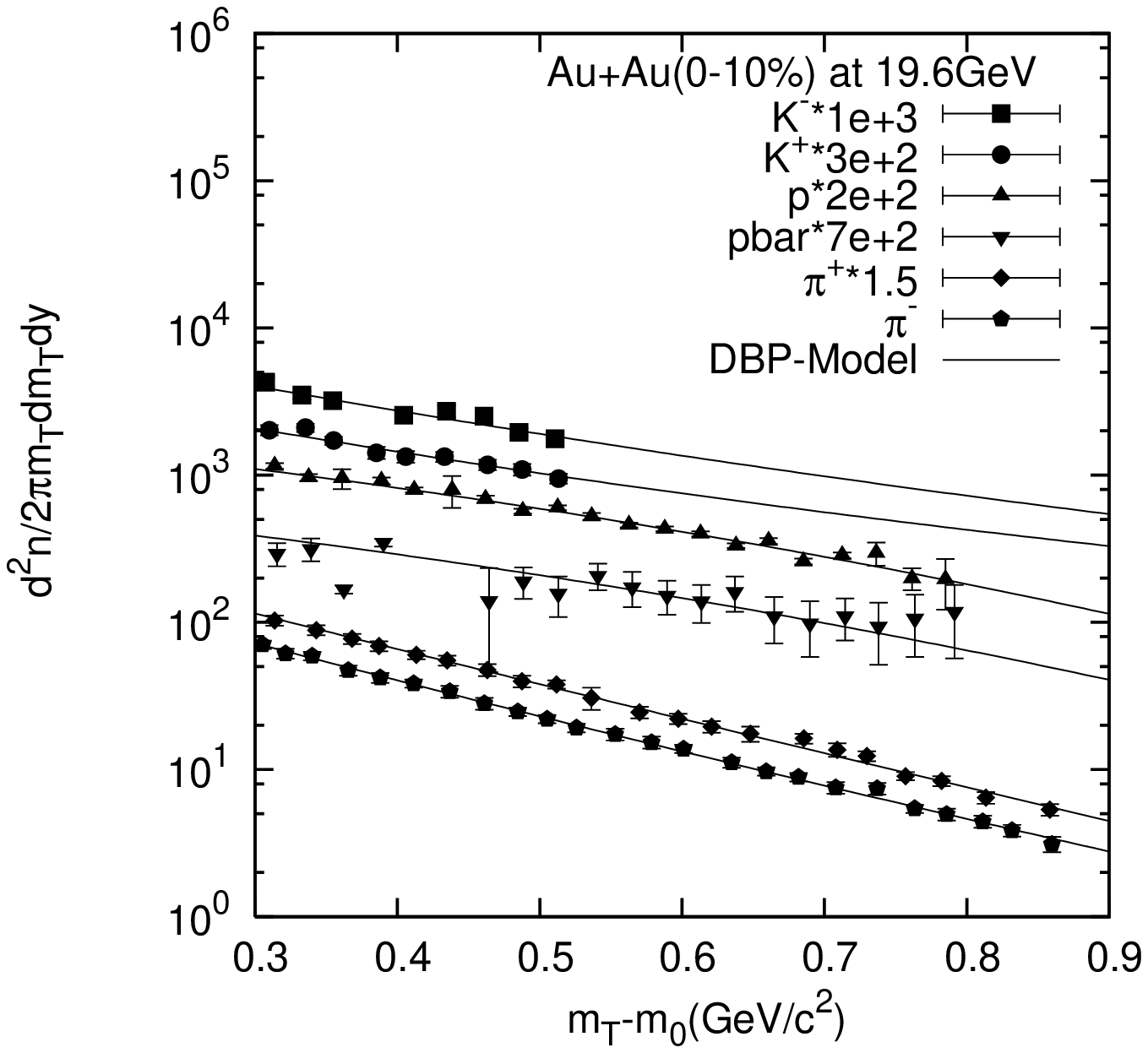}
\setcaptionwidth{2.6in}
\end{minipage}}%
\subfigure[]{
\begin{minipage}{0.5\textwidth}
\centering
 \includegraphics[width=2.5in]{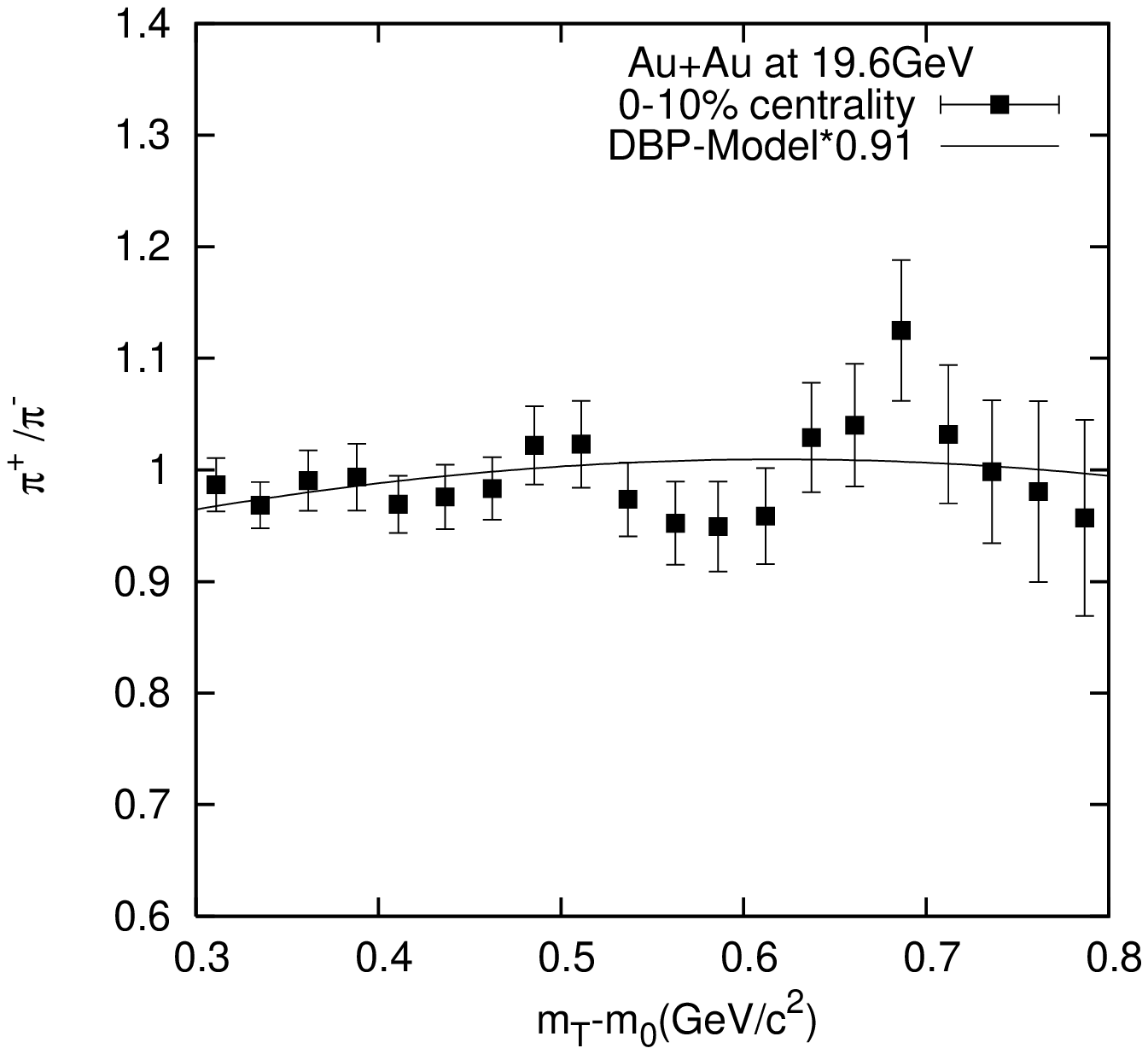}
 \end{minipage}}%
\vspace{0.01in} \subfigure[]{
\begin{minipage}{0.5\textwidth}
\centering
\includegraphics[width=2.5in]{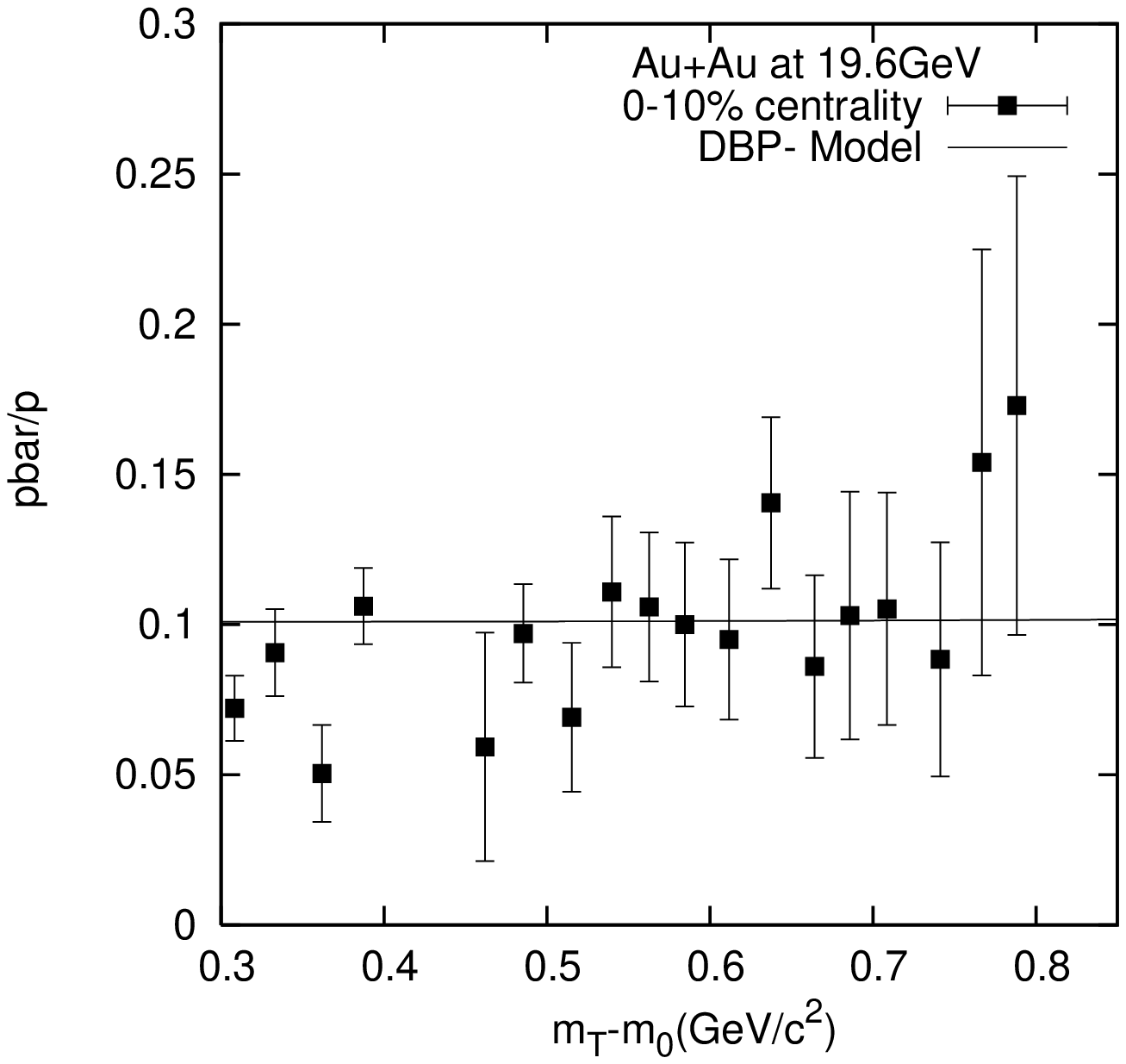}
\end{minipage}}%
\subfigure[]{
\begin{minipage}{.5\textwidth}
\centering
 \includegraphics[width=2.5in]{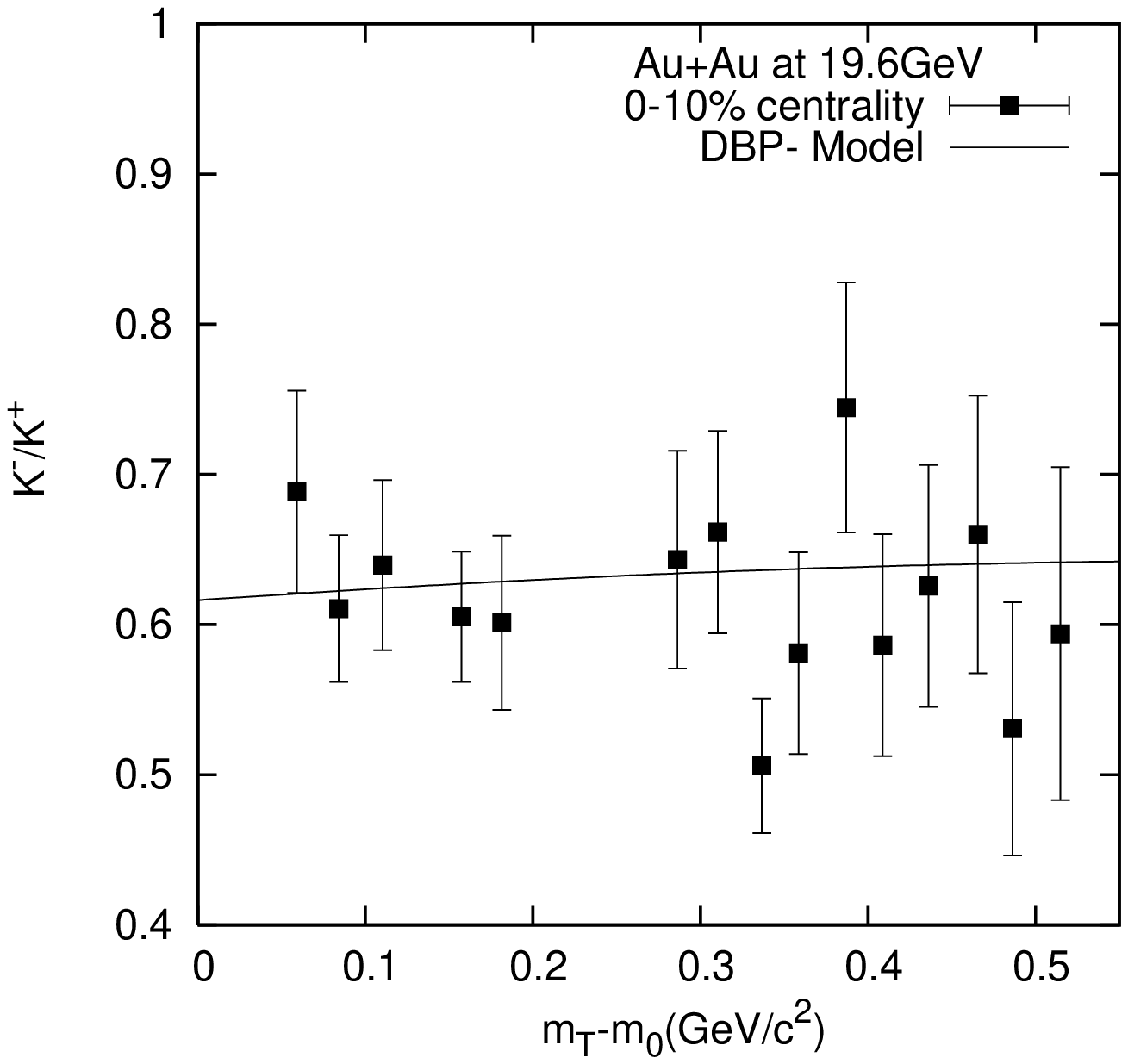}
 \end{minipage}}%
\caption{(a) Transverse mass spectra of identified hadrons measured at midrapidity $(|y|<0.1)$. The results at
$\sqrt{s_{NN}}$=19.6 GeV for the production of $\pi^+$, $\pi^-$, p, $\overline{p}$, $K^+$ and $K^-$
for 0-10\% centrality in Au+Au collisions. The solid curves provide the DBP Model based results. Data are taken from Ref.\cite{Picha1}
(b),(c),(d) $\pi^-/\pi^+$, $\overline{p}/p$ and $K^-/K^+$ ratios vs. $m_T-m_0$ for 0-10\% centrality in Au+Au collisions at
 $\sqrt{s_{NN}}$=19.6 GeV $(-0.1<y<0.1)$. The solid curves provide the DBP Model based results. Data are taken from Ref.\cite{Picha1}
 and all errors are only of statistical nature.}
\end{figure}

\begin{figure}
\subfigure[]{
\begin{minipage}{.5\textwidth}
\centering
\includegraphics[width=2.5in]{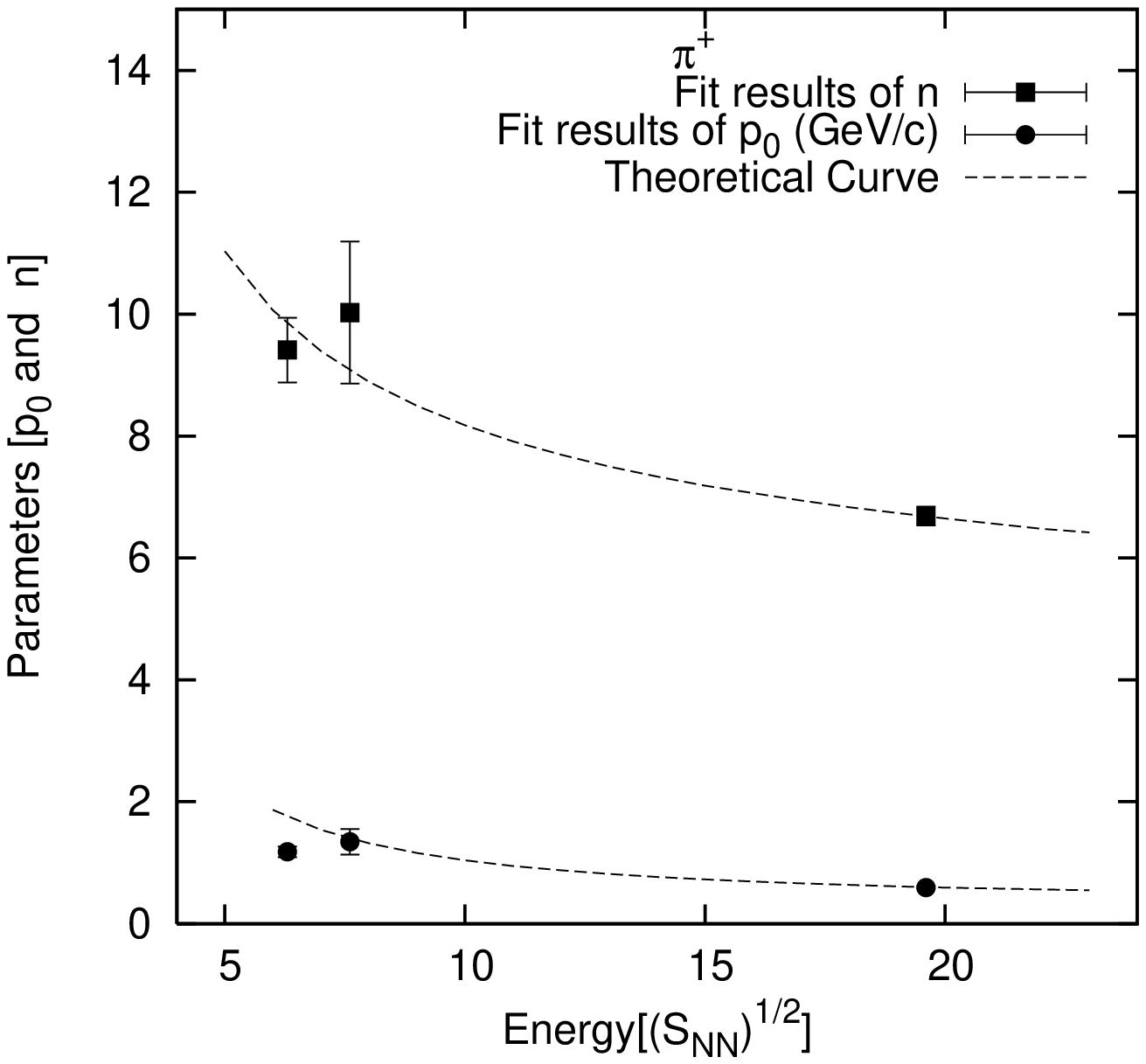}
\end{minipage}}%
\subfigure[]{
\begin{minipage}{.5\textwidth}
\centering
 \includegraphics[width=2.5in]{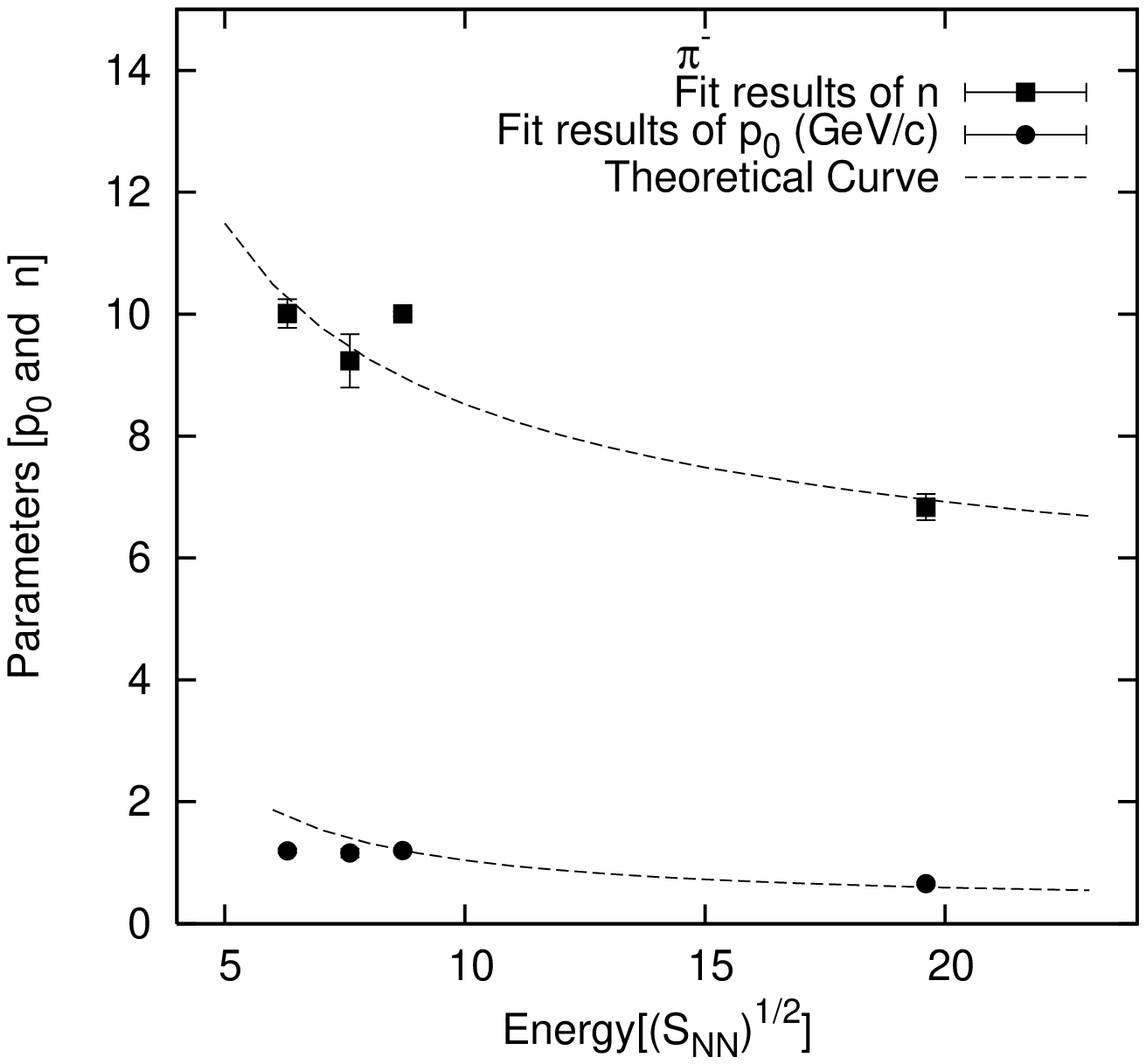}
 \end{minipage}}%
\caption{Values of $p_0$ and n as a function of c.m. energy $\sqrt{S_{NN}}$. The dotted curves are drawn for
$\pi^+$ and $\pi^-$ on the basis of eqn. (11) and (12). }
\end{figure}

\begin{figure}
\subfigure[]{
\begin{minipage}{.5\textwidth}
\centering
\includegraphics[width=2.5in]{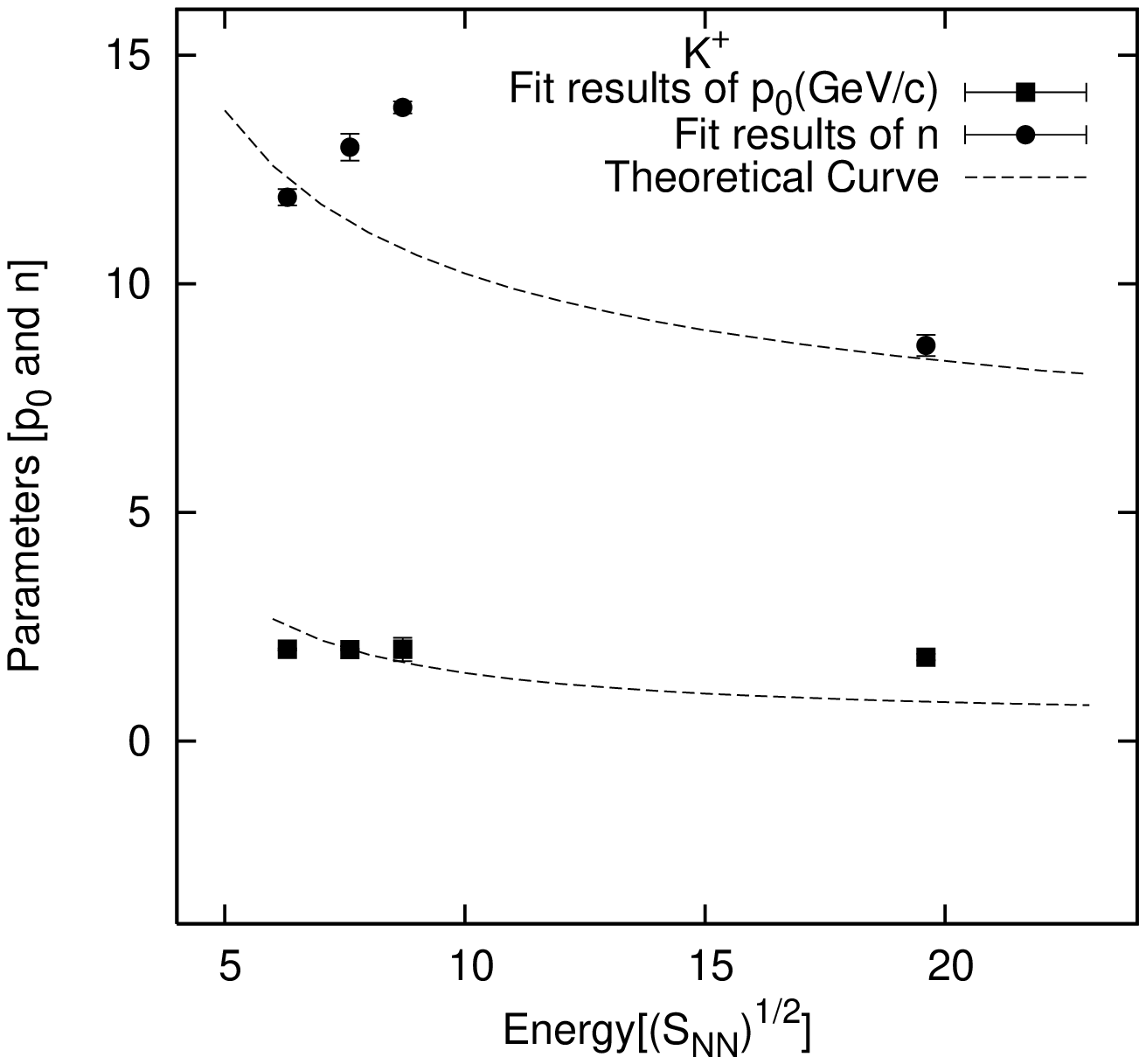}
\end{minipage}}%
\subfigure[]{
\begin{minipage}{.5\textwidth}
\centering
 \includegraphics[width=2.5in]{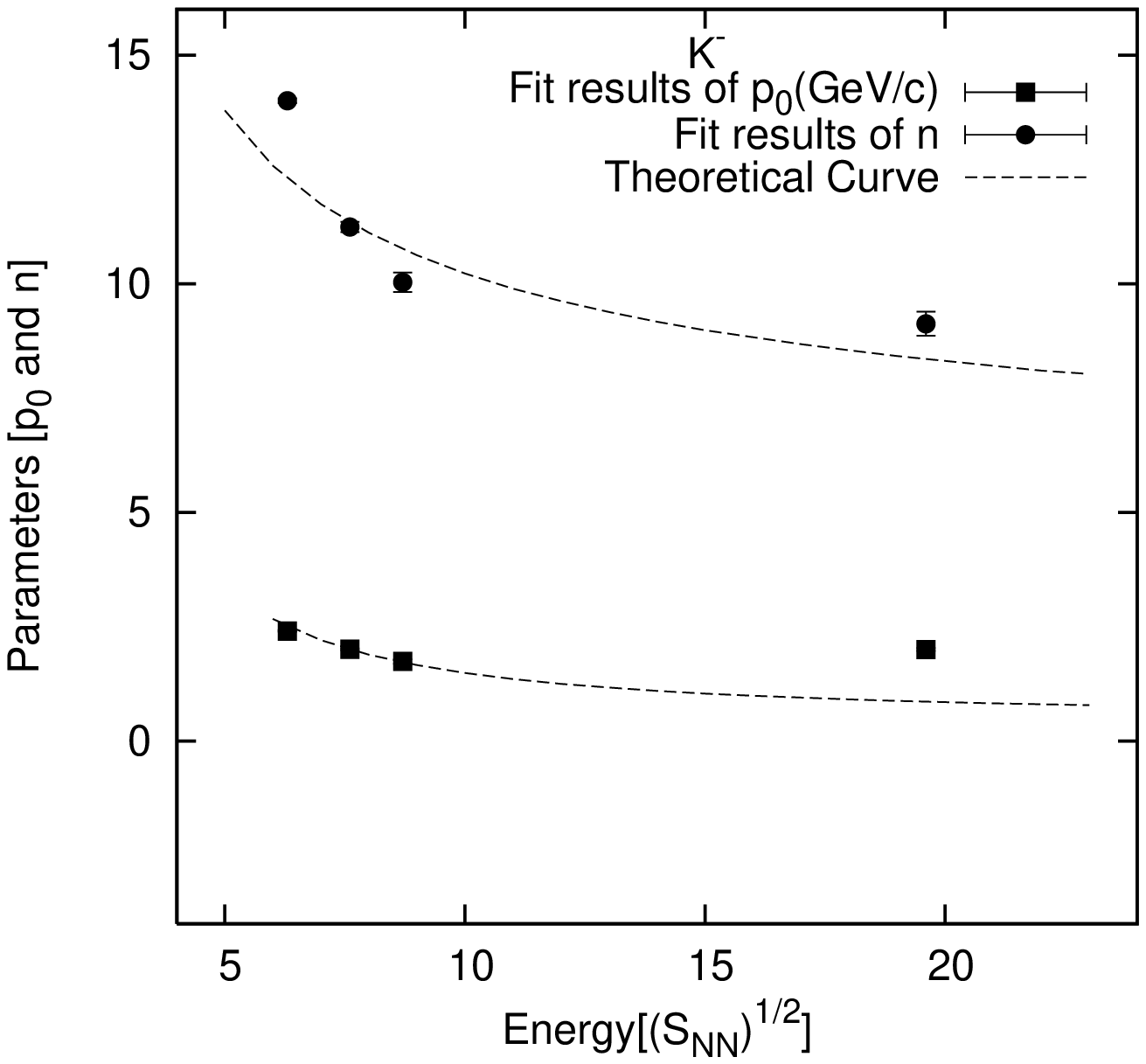}
 \end{minipage}}%
\caption{Values of $p_0$ and n as a function of c.m. energy $\sqrt{S_{NN}}$. The dotted curves are drawn for
$K^+$ and $K^-$ on the basis of eqn. (11) and (12). }
\end{figure}

\begin{figure}
\subfigure[]{
\begin{minipage}{.5\textwidth}
\centering
\includegraphics[width=2.5in]{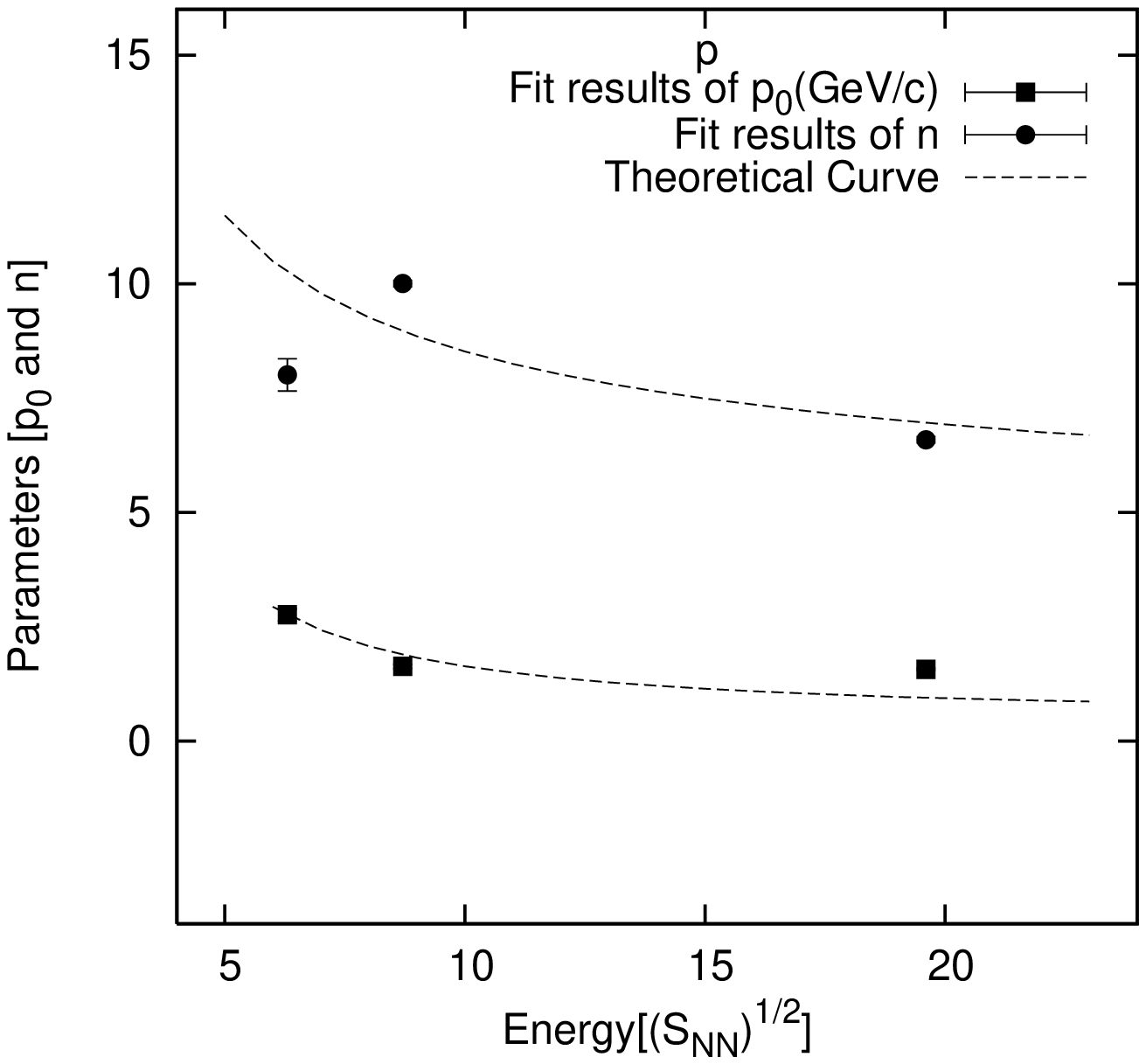}
\end{minipage}}%
\subfigure[]{
\begin{minipage}{.5\textwidth}
\centering
 \includegraphics[width=2.5in]{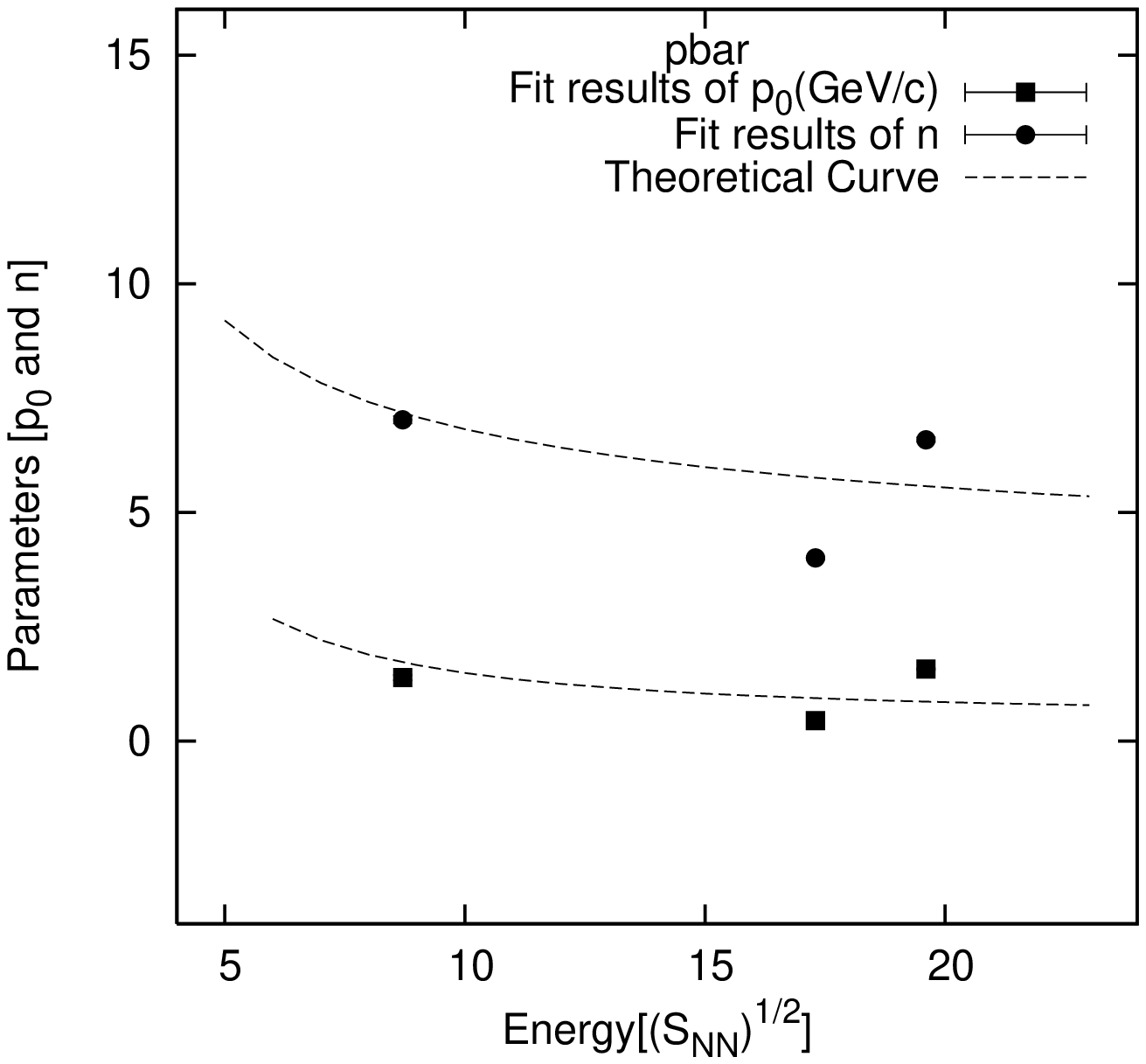}
 \end{minipage}}%
\caption{Values of $p_0$ and n as a function of c.m. energy $\sqrt{S_{NN}}$. The dotted curves are drawn for
$p$ and $\overline{p}$ on the basis of eqn. (11) and (12). }
\end{figure}


\begin{thebibliography}{*}
\bibitem{Mukhi1} Sunil Mukhi and Probir Roy : Pramana {\bf 73},
(2009), 3-60.
\bibitem{Busza1} Wit Busza : Nucl. Phys. {\bf A 830}, (2009), 35c-42c.
\bibitem{Back1}B.B.Back : Phys. Rev. {\bf C 72}, (2005), 064906 [nucl-ex/0508018 v1 15 August 2005].
\bibitem{De1} B.De, S.Bhattacharyya and P.Guptaroy : J. Phys. {\bf G28}, (2002), 2963.
\bibitem{De3}B.De and S.Bhattacharyya : Int. J. Mod. Phys. {\bf A 19}, (2004), 3225.
\bibitem{De4}B.De and S.Bhattacharyya : Mod. Phys. Lett. {\bf A 18}, (2003), 1383.
\bibitem{De2}B.De and S.Bhattacharyya : Eur. Phys. J. {\bf A 19}, (2004), 237.
\bibitem{Arnison1}UA1 Collaboration(G.Arnison et al.) : Phys. Lett. {\bf B 118}, (1982), 167.
\bibitem{Bocqet1}UA1 Collaboration(G.Bocqet et al.) : Phys. Lett. {\bf B 366}, (1996), 434.
\bibitem{Albrecht1}WA80 Collaboration(R.Albrecht et al.) : Eur. J. {\bf C 5},
(1998),255.
\bibitem{Hagedorn1}R.Hagedorn : Riv. Nuovo. Cimento 6, (1983), 46; R.Hagedorn: CERN-TH.3684,1983.
\bibitem{Peitzmann1}T.Peitzmann : Phys. Lett. {\bf B 450}, (1999), 7 .
\bibitem{Arsene1}BRAHMS Collaboration(I.Arsene et al) : Phys. Rev. Lett. {\bf 91}, (2003), 072305.
\bibitem{Sau1}G. Sau, S. K. Biswas, A. C. Das Ghosh, A. Bhattachrya
and S. Bhattacharyya : IL Nuovo Cimento {\bf B 125}, (2010), 833.
\bibitem{Lehmann1}S. Lehmann, A. D. Jackson and B. E. Lantrap :
Physics/0512238 v1 24 December 2005.
\bibitem{Sarcevic1}Ina Sarcevic : Proc. of Large Hadron Collidier Workshop
vol.II, Edited by G. Jarlskog and D. Rein : CERN 90-10, ECFA
10-133(03 December 1990) 1214-1223.
\bibitem{Biro1}T. S. Biro and G. Purcsel : Phys. Rev. Lett {\bf 95},
(2005), 162302.
\bibitem{Biswas1}S. K. Biswas, G. Sau, B. De, A. Bhattacharya and S.
Bhattacharyya : Hadronic Journal {\bf 30}, (2007), 533-554.
\bibitem{Williams1} Mike Williams : hep-ex/1006.3019 v1 15 June 2010.
\bibitem{Tsallis1}C. Tsallis : J. Stat. Phys. {\bf 52}, (1988), 479; Physica
{\bf A 221}, (1995), 277; Braz. J. Phys. {\bf 29}, (1999), 1; D.
Prato and C. Tsallis : Phys. Rev. {\bf E 60}, (1999), 2398.
\bibitem{Alt1}NA49 Collaboration (C.Alt et al) : Phys. Rev. {\bf C 77}, (2008), 024903 [nucl-ex/0710.0118 v1
30 September 2007].
\bibitem{Alt2}NA49 Collaboration (C.Alt et al) : Phys. Rev. {\bf C 77}, (2008), 064908 [nucl-ex/0709.4507 v2
26 September 2007].
\bibitem{Alt3}NA49 Collaboration (C.Alt et al) : Phys. Rev. {\bf C 77}, (2008), 034906  [nucl-ex/0711.0547 v2 28 March 2008].
\bibitem{Cebra1}STAR Collaboration (D.Cebra) : nucl-ex/0903.4702 v1 (Submitted on 26 March 2009).
\bibitem{Picha1}Roppon Picha : Ph. D. Thesis, University of California, Davis, USA
(2005).

\end{thebibliography}
\end{document}